\documentclass[a4paper,fleqn,usenatbib]{mnras}

\usepackage{newtxtext,newtxmath}
\usepackage{mathptmx}
\usepackage{txfonts}

\usepackage[T1]{fontenc}
\usepackage{ae,aecompl}

\usepackage{lipsum}

\usepackage{graphicx}	
\usepackage{amsmath}	
\usepackage{amssymb}	
\usepackage{booktabs}
\usepackage{multirow}

\newcommand\E{1E{\thinspace}1740.7$-$2942}

\title[Broadband X-ray analysis of \E]{Broadband X-ray analysis of \E: constraints on spin, inclination and a tentative black hole mass}

\author[Stecchini et al.]{
Paulo E. Stecchini$^{1}$\thanks{E-mail: paulo.stecchini@inpe.br},
F. D'Amico$^{1}$,
F. Jablonski$^{1}$,
M. Castro$^{2}$
and J. Braga$^{1}$
\\
$^{1}$Instituto Nacional de Pesquisas Espaciais, Av. dos Astronautas 1758, 12227-010, S.J. Campos--SP, Brazil\\ 
$^{2}$Universidade Estadual de Campinas, Cidade Universitaria Zeferino Vaz, 13083-853, Campinas--SP, Brazil
}

\date{Accepted XXX. Received YYY; in original form ZZZ}

\pubyear{2020}

\begin{document}
\label{firstpage}
\pagerange{\pageref{firstpage}--\pageref{lastpage}}
\maketitle

\begin{abstract}
\E\space is one of the strongest hard X-ray emitters in the Galactic Centre region, believed to be a black hole in a high-mass X-ray binary system. Although extensively studied in X-rays, many aspects about the underlying nature of the system are still unknown. For example, X-ray data analyses of \E\space up to date have not yet unveiled the signature of a reflection component, whose modelling could be used to estimate parameters such as the spin of the black hole and inclination of the disc. We report here on the determination of these parameters from the analysis of the reflection component present in a public \textit{NuSTAR} observation which hasn't been subject to any previous study. We include \textit{XMM-Newton} and \textit{INTEGRAL} data to build a combined spectrum, enabling a joint analysis of both the disc and comptonisation components. Results point to a relatively high inclination disc $\gtrsim$\,50$^{\circ}$ (3\,$\sigma$) and a near-maximum speed rotating black hole. The former is in agreement with a  previous radio study and the latter is reported here for the first time. Lastly, we follow the methodology of recent efforts to weight black holes with only X-ray spectra and find results that suggest a black hole mass of about 5\,M$_\odot$ for \E.

\end{abstract}




\section{Introduction}

The source \E\space was discovered by the \textit{Einstein Observatory} satellite during the first Galactic Plane X-ray survey \citep{1984ApJ...278..137H} and classified as a black hole candidate due to its spectral similarities to Cygnus X-1 \citep{1991ApJ...383L..49S}. This classification was later supported by the observation of radio jets coming from the main X-ray source \citep{1992Natur.358..215M}, which also identified \E\space as a microquasar.  Known to be among the brightest hard X-ray sources around the Galactic Centre (GC), \E\space is observed to spend most of the time in the low/hard state (LHS) of emission (see, e.g., \citealp{cas141}) -- a state well described by thermal comptonisation models. The permanence in this state had already been verified from studies with data from \textit{RXTE}\,+\,\textit{INTEGRAL} \citep{2005A&A...433..613D}, the telescopes on-board \textit{Suzaku} \citep{2010AIPC.1248..189R} and \textit{NuSTAR}\,+\,\textit{INTEGRAL} \citep{2014ApJ...780...63N}. These last two works reported no strong evidence of a reflection component in their spectra. There is, until now, no optical or infrared counterpart confirmed for \E\space -- which is usually attributed to the high galactic extinction towards the GC at these wavelengths (see, e.g., \citealp{2002MNRAS.337..869G}). In one of the most recent attempts, \cite{2010ApJ...721L.126M} have successfully identified a single near-infrared source towards the system's location. The study suggested two hypotheses to explain the counterpart candidate detected: an Active Galactic Nucleus projected along the line of sight of the system or a black hole high-mass X-ray binary. The former hypothesis was discarded by \cite{2015A&A...584A.122L} after analysing the bi-polar radio jets of \E. A periodic modulation of $\sim$\,12.6 days, reported by \cite{2002ApJ...578L.129S} and \cite{2017ApJ...843L..10S}, if attributed to the orbital period of the system, also supports the proposition of \E\ being part of a high-mass X-ray binary system, as this modulation is much longer than most known black hole low-mass X-ray binaries orbital   periods (see, e.g., \citealp{2016A&A...587A..61C}). Without an observable counterpart the relative motions between the compact object and the companion cannot be determined, preventing the mass function $f$($M$) of the system from being established. Hence, apart from the rough estimate of the distance to the source -- assumed to be that to the GC -- and the suggestion of a high inclination disc (from the presence of bi-polar radio jets, \citealp{1992Natur.358..215M}), many dynamical parameters are still unknown for \E. 

In this study we gather public available data of \E\space from \textit{XMM-Newton} \citep{2001A&A...365L...1J}, \textit{NuSTAR} \citep{2013ApJ...770..103H} and \textit{INTEGRAL} \citep{2003A&A...411L...1W} to build a broadband spectrum, whose total energy coverage ($\sim$\,2--200\,keV) allows us to inspect the disc, reflection and comptonisation components altogether. Using the X-ray spectral-fitting program \textsc{xspec} \citep{1996ASPC..101...17A} we apply from simple powerlaw models to some of the most popular comptonisation models available and compare our results with values found in the literature for \E. Our main goal, however, is to model the reflection component, which we regard to be present in the \textit{NuSTAR} data. The presence of this component is a necessary condition to apply the so-called X-ray reflection spectroscopy method (or iron line method), which can directly provide parameters such as the spin and inclination of the system. 
Finally, motivated by the lack of information with respect to the mass of the system and by the recent efforts to weight black holes from X-ray spectra alone (see, e.g., \citealp{2016ApJ...821L...6P}), we analyse the disc spectrum (which, due to \textit{XMM-Newton} data we can model down to 2\,keV) with a mass-dependent model to speculate over this parameter.

\section{Observations}

\subsection{\textit{NuSTAR}}

\E\space was observed for calibration purposes by the \textit{NuSTAR} satellite on three occasions, all of which occurred  within $\sim$\,2 months of the mission's launching. Two of the observations (of $\sim$2\,ks and $\sim$6\,ks exposure times) have been analysed by \cite{2014ApJ...780...63N}, which reported the absence of any strong  iron line or reflection components. We focus our analysis on the other observation -- the longest available (of $\sim$\,10\,ks) -- and which hasn't yet been subject to any study found in the literature. The data from the observation (ObsID 10002012001) were reduced using the standard procedures of the \textit{NuSTAR} Data Analysis Software (NuSTARDAS) for both FPMA and FPMB cameras. Source and background spectra were extracted from a $\sim$\,90$''$ circular region centred on the source position and from a source-free region of $\sim$\,150$''$, respectively. Due to a possible overwriting issue during the observation, FPMB camera had $\sim$\,20$\%$ less exposure time than FPMA and therefore was not included in this analysis. The FPMA data were rebinned to have at least 50 counts per spectral bin prior to the analysis conducted with \textsc{xspec}. 

\begin{figure}
	\includegraphics[width=\columnwidth]{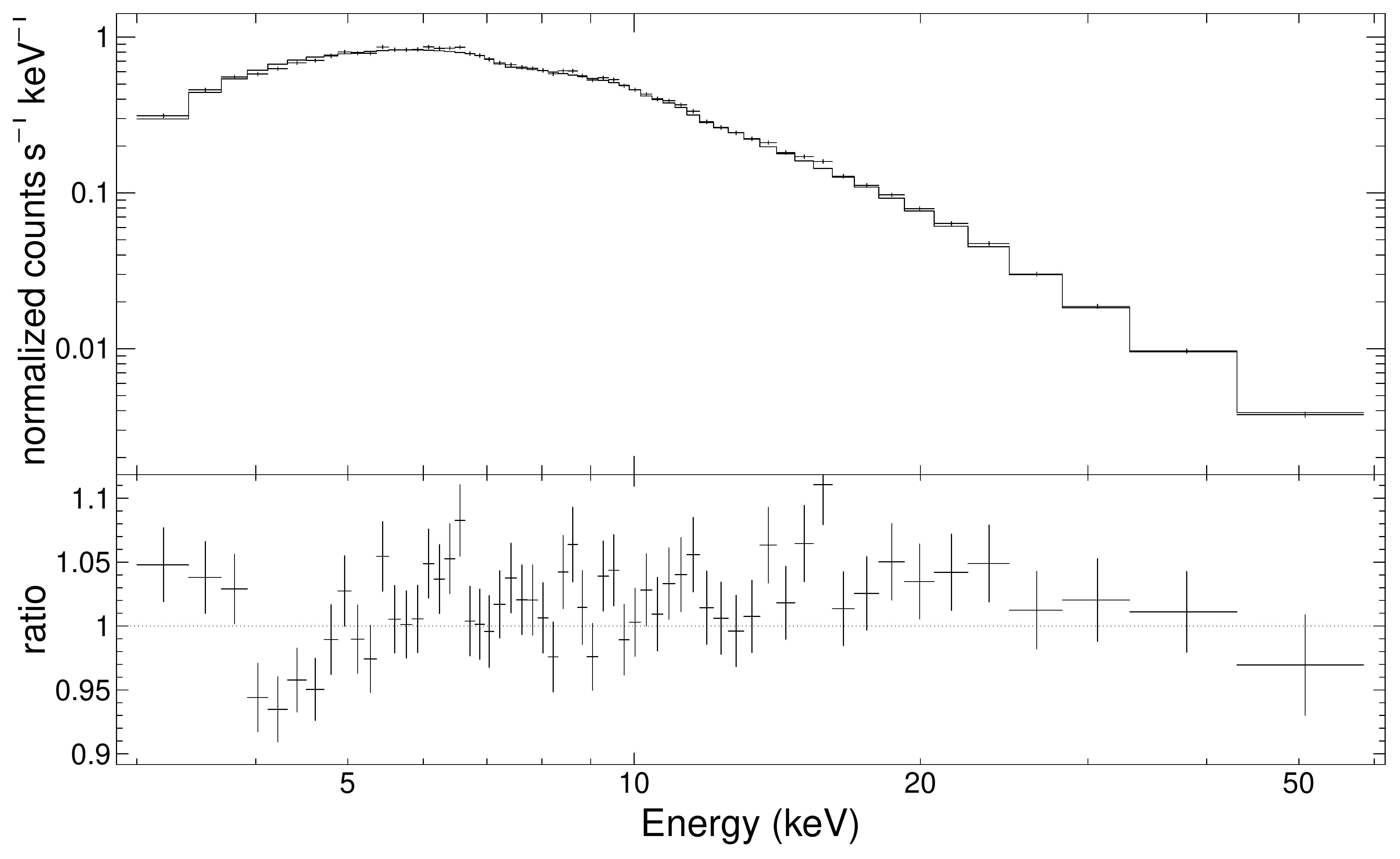}
    \caption{Spectrum of \E\space from FPMA and the residuals for an absorbed powerlaw model. Data were rebinned for plotting purposes only.}
    \label{fig:01}
\end{figure}

Figure \ref{fig:01} shows (rebinned for plotting purposes only) the data and the residuals (as $\text{data/model}$) for an absorbed powerlaw model (\texttt{phabs*powerlaw} from \textsc{xspec}). The best-fitting ($\chi^2/\nu$\,=\,$\frac{562}{507}$\,=\,1.11, where $\chi^2$ is the usual chi-square statistics, $\nu$ is the number of degrees of freedom and the last value is the reduced $\chi^2$) parameters were $N_H$ ($\times$\,10$^{22}$)\,=\,13.2\,$\pm$\,1.5\,cm$^{-2}$ and a powerlaw index ($\Gamma$) of 1.76\,$\pm$\,0.03. These values differ slightly from those reported for the other two \textit{NuSTAR} observations (i.e., $N_H$ > 18 $\times$\,10$^{22}$ and $\Gamma \lesssim$\,1.7); the interstellar absorption found here is in better agreement with previously reported values from other missions (see, e.g., \citealp{2002MNRAS.337..869G}, \citealp{2010AIPC.1248..189R} and \citealp{cas141}) and the steeper photon index -- although within the range of indices that describe well the low/hard state (1.4\,$\leq$ $\Gamma$\,$\leq$\,2.1, see, e.g., \citealp{2006ARA&A..44...49R}) -- may indicate that \E\space was going through a state transition or was in a ``softer'' low/hard state. As also pointed out by \cite{2014ApJ...780...63N} for the other two observations, the residuals at lower energies suggest that a soft component is required. In fact, adding a multicolour disc blackbody component (\texttt{diskbb}, \citealp{1984PASJ...36..741M}; \citealp{1986ApJ...308..635M}) to the model and applying an F-test gives a very low probability ($\sim$\,10$^{-6}$) against the null hypothesis.
Also noticeable from the residuals in Figure \ref{fig:01} are a slight excess at about 15--30\,keV -- which could be due to a reflection feature -- and a potential emission line between 6--7\,keV. Expecting the latter to be a broad Fe K$\alpha$ line, we include a gaussian initially centred at 6.4\,keV to the model. The best fit places it at 6.44\,$\pm$\,0.15\,keV and the calculated upper limits for the equivalent widths (EW) for different lines with full widths at half maximum (FWHM) of 0.25, 0.5 and 1.0\,keV are 49, 58 and 72\,eV, respectively. A simple test is also performed to verify the likelihood of the presence of the 6.4\,keV feature in the data. We compare the expected counts from the smooth powerlaw continuum in the region with that of a possible feature superimposed on that continuum and spread along four energy bins centred at 6.08, 6.24, 6.4 and 6.56\,keV. The $\chi^2$ statistics can be defined as
\begin{equation}
\chi^{2}_4\text{=}\,  \frac{\sum_{i\text{=}1}^4 \text{(}\text{data}_i - \text{model}_i\text{)}^2}{\sigma^2},
\end{equation}

\noindent where data$_i$ and model$_i$ refer to the four bins closer to the 6.4\,keV feature and $\sigma$, the same for all four bins, is estimated from the four energy bins just before and after the expected 6.4\,keV line. The $\chi^2_4$ for the real data gives a value of 41.54, as opposed to a median $\chi^2_4$ of 3.35 for 10$^6$ randomly generated sets of four bins. This suggests that a line feature is required to better describe the spectrum in this region with a confidence of $1:10^{8}$ parts (roughly 5.7\,$\sigma$ in a normal distribution).

\subsection{\textit{XMM} and \textit{INTEGRAL}}

Based on the individual powerlaw index that best fitted the \textit{NuSTAR} observation, we chose two observations -- from PN, on-board \textit{XMM-Newton} (ObsID 0303210201) and from ISGRI, on-board \textit{INTEGRAL} (Revolution 1200) -- to build a broadband spectrum. Data from these instruments were reduced using each mission's standard procedures: The \textit{XMM-Newton} Science Analysis System (SAS) for PN and \textit{INTEGRAL} Off-line Scientific Analysis (OSA) for ISGRI. A summary of the observations (including \textit{NuSTAR}) with the date, exposure time and the individual best-fitting powerlaw indices is shown in Table \ref{tab:01}. It can be noticed that the observations were not contemporary; \textit{XMM} and \textit{NuSTAR} observations are separated by almost 7 years. 
The primary argument for a simultaneous fit is the very similar photon indices (last column of Table \ref{tab:01}), suggesting that \E\space was in the same spectral state on the three occasions.

\begin{table}
	\centering
	\caption{Summary of the observations with the date, exposure time and the individual best-fitting powerlaw indices.}
	\resizebox{\columnwidth}{!}{%
	\hspace{-0.55cm}
	\label{tab:01}
	\begin{tabular}{lccc} 
\multicolumn{1}{c}{Telescope} & Observation Date & Exposure Time (s) & $\Gamma$ \\ \hline
XMM - PN & 02/10/2005 & 16,040 & 1.77\,$\pm$\,0.04 \\
NuSTAR - FPMA & 03/07/2012 & 10,970 & 1.76\,$\pm$\,0.03 \\
INTEGRAL - ISGRI & 13/08/2012 & 9,332 & 1.79\,$\pm$\,0.10 \\ \hline
	\end{tabular}
	}
\end{table}

\begin{figure}
	\includegraphics[width=\columnwidth]{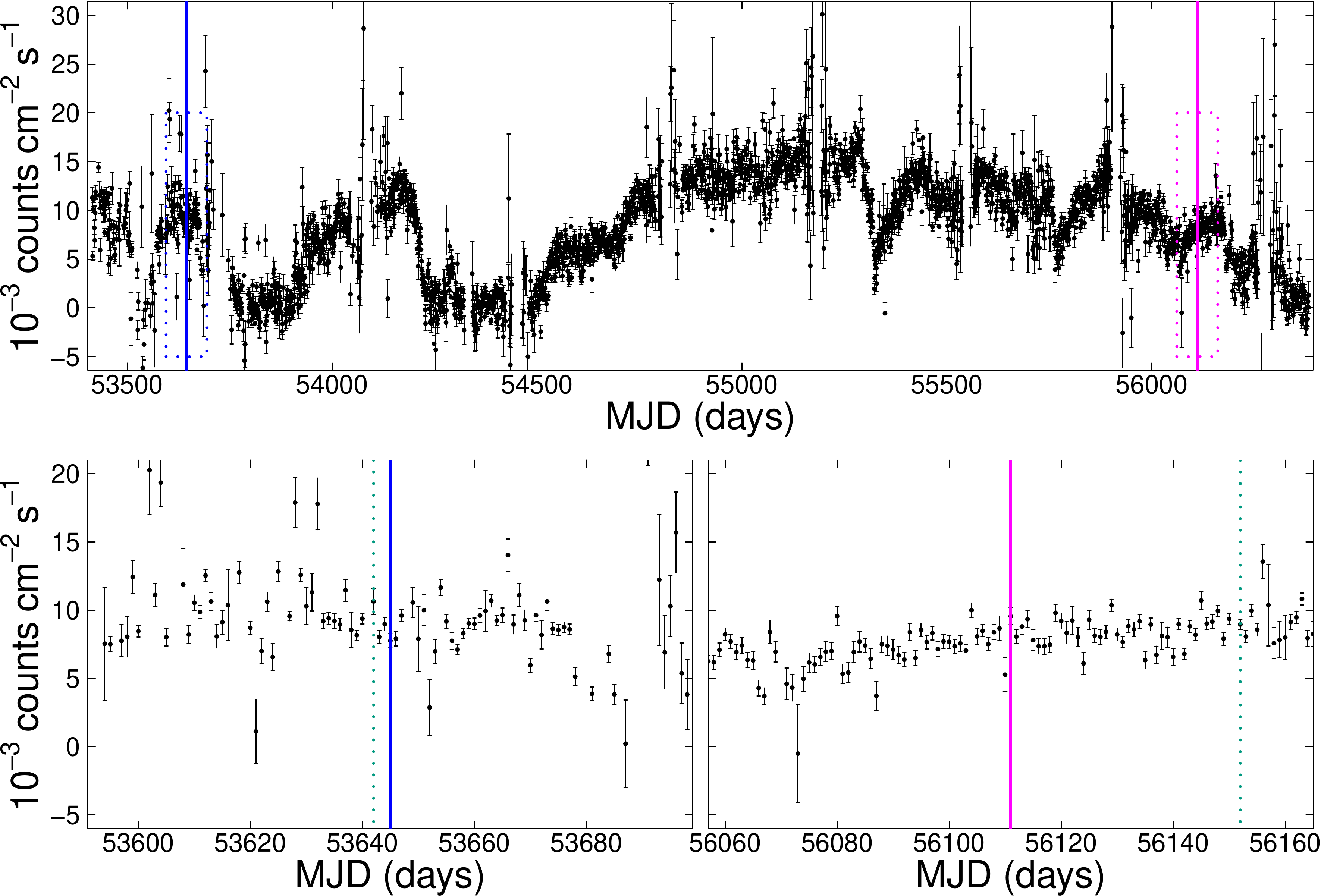}
    \caption{\textit{Upper panel}: Swift BAT long--term light curve for \E. The blue and magenta lines show XMM-Newton and NuSTAR observation dates respectively. \textit{Lower panel}: A zoom of the region delimited by the dotted boxes drawn in the full curve ($\pm$50\,days from the observations dates). The bluish green dotted lines show the closest INTEGRAL observations.}
    \label{fig:02}
\end{figure}

Additional support for combining the spectra comes from the long--term light curve provided by the Burst Alert Telescope (BAT) on-board the \textit{Swift}  satellite \citep{2004ApJ...611.1005G} which allows us to examine and compare the flux of \E\space during the epochs it was observed by \textit{XMM} and \textit{NuSTAR}.
The BAT instrument has been monitoring  the sky since $\sim$\,2005 and provides daily count rates for many astrophysical sources in the 15--50\,keV band. The upper panel of Figure \ref{fig:02} displays the daily flux for \E\ from 2005 February 14 (MJD 53415) to 2013 April 1 (MJD 56383). The observation dates of \textit{XMM} (MJD 53645) and \textit{NuSTAR} (MJD 56111) are indicated by the blue and magenta solid lines, respectively; the dotted boxes with the same colours delimit 50 days before and after these dates, which are displayed in detail in the lower panel. 
During the whole period, the median flux of \E\space in the 15--50\,keV band was 8.9\,$\times$\,10$^{-3}$\,counts\,$\cdot$\,cm$^{-2}$\,$\cdot$\,s$^{-1}$ with a standard deviation of 5.4\,$\times$\,10$^{-3}$\,counts\,$\cdot$\,cm$^{-2}$\,$\cdot$\,s$^{-1}$, whilst the flux for \textit{XMM} and \textit{NuSTAR} observation were 7.7$\pm$1.0\,$\times$\,10$^{-3}$ and 9.0$\pm$0.8\,$\times$\,10$^{-3}$\,counts\,$\cdot$\,cm$^{-2}$\,$\cdot$\,s$^{-1}$, respectively. These values express that, despite the long gap between the observations, their fluxes are not only consistent with the total median flux (well within 1\,$\sigma$) but are also marginally comparable to each other considering 1\,$\sigma$ error bars in each dataset.

\begin{figure}
	\includegraphics[width=\columnwidth]{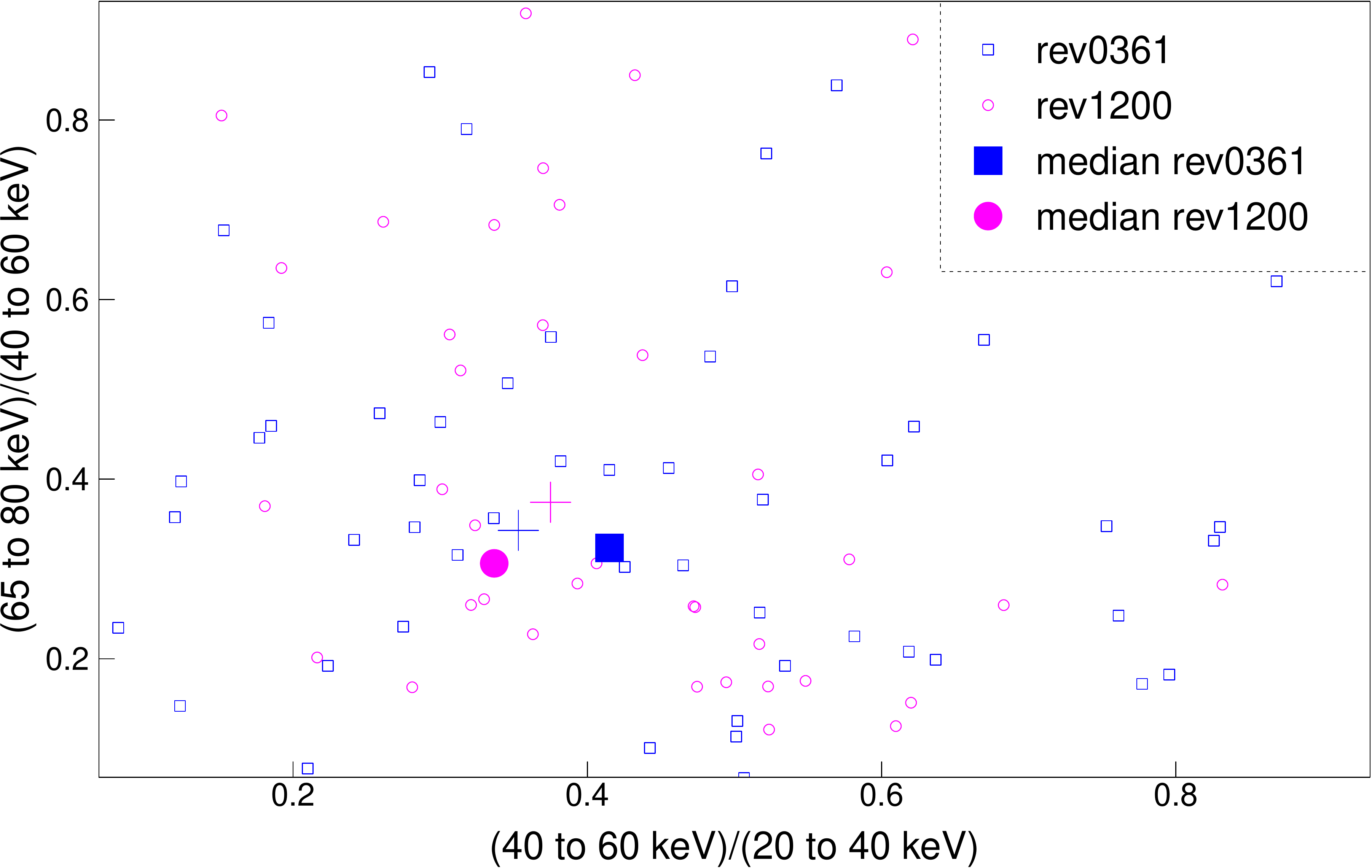}
    \caption{Colour-colour diagram (unfilled shapes) for the INTEGRAL observations whose dates were shown in Figure \ref{fig:02}. The medians of all the colour ratios for each observation are shown as the slightly bigger filled shapes. The flux ratios for the same bands are also shown (as the plus symbols).}
    \label{fig:03}
\end{figure}

The bluish green dotted lines in the lower panel of Figure \ref{fig:02} point out two near \textit{INTEGRAL} ISGRI observations in time to the \textit{XMM} (Revolution 0361) and to the \textit{NuSTAR} (Revolution 1200) observations. As a supplementary way to verify the source's behaviour during both periods, we present in Figure \ref{fig:03} a colour-colour diagram for these ISGRI observations. The diagram is built, for each observation, by extracting light curves for three different energy bands -- namely a soft (defined here as 20 to 40\,keV), a medium (40 to 60\,keV) and a hard band (65 to 80\,keV) -- and plotting the ratio $hard/medium$ bands against the ratio $medium/soft$ bands. In Figure \ref{fig:03}, the open symbols represent the colours of ISGRI observations closest to \textit{XMM} (squares) and \textit{NuSTAR} (circles) and the slightly larger filled symbols are their corresponding median values. Also shown (as plus symbols) are the flux ratios -- for the same energy bands mentioned -- calculated for each ISGRI spectra after a powerlaw fit. Their proximity in the diagram indicates that the source was indeed in a very similar state in both occasions. Due to the overlapping energy band, we choose to use the ISGRI observation closer to that of \textit{NuSTAR} (Revolution 1200) to compose the higher energy end of the spectrum.

\section{Broadband Analysis and Results}

\begin{figure}
	\includegraphics[width=\columnwidth]{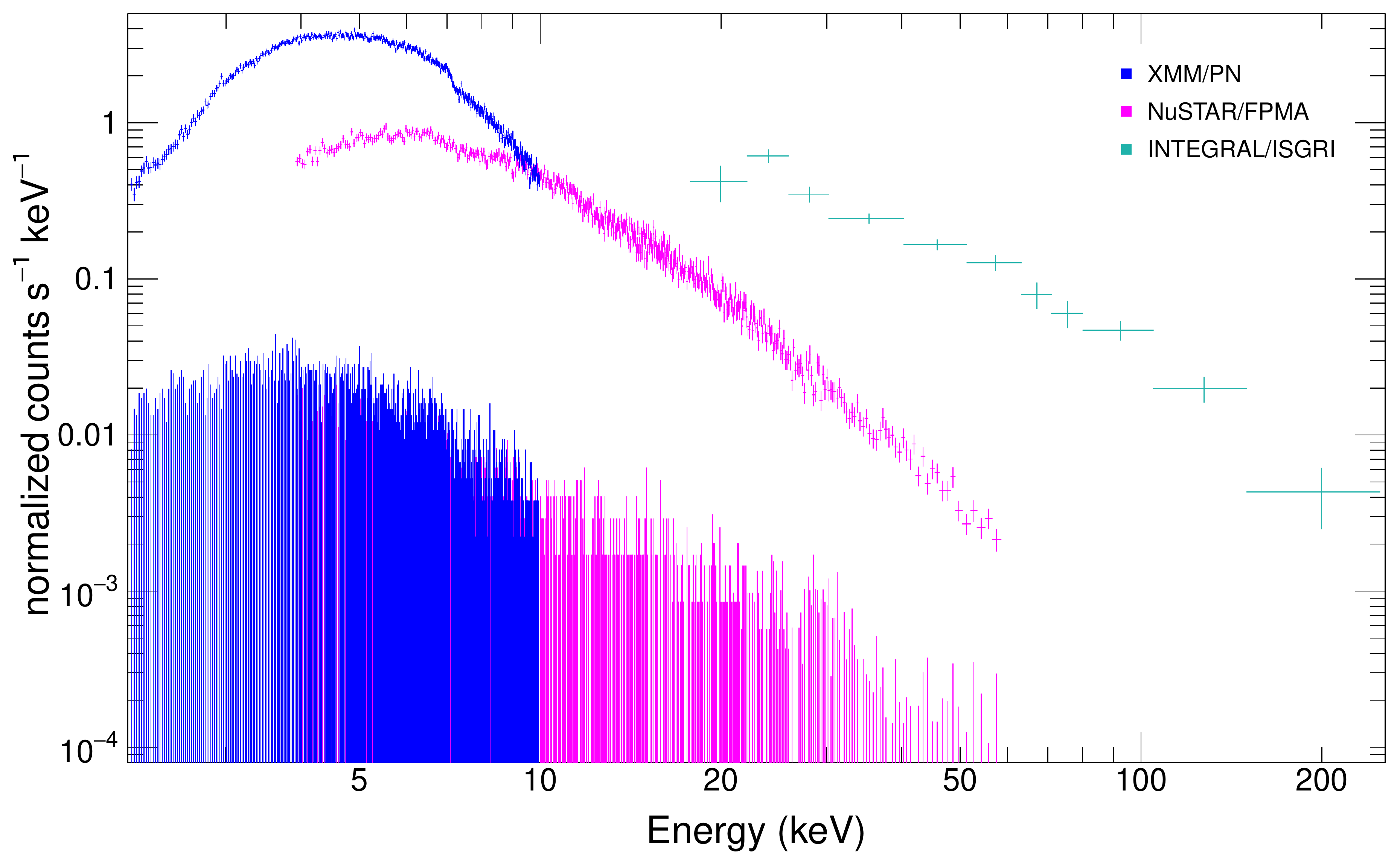}
    \caption{Composed spectrum and background of \E\space with XMM/PN (blue), NuSTAR/FPMA (magenta) and INTEGRAL/ISGRI (bluish green) spectra.}
    \label{fig:04}
\end{figure}

Figure \ref{fig:04} shows the combined spectrum taking into account the considerations from the previous section. The following energy bounds are set for each instrument: \textit{XMM-Newton}: 2--10\,keV; \textit{NuSTAR}: 4--60\,keV and \textit{INTEGRAL}: 20--200\,keV. All models to be described include a multiplicative \texttt{constant} to correct the flux offset and all quoted errors are at 90\% confidence unless indicated otherwise.

\subsection{Powerlaw Models}

We initiate the composed spectrum analysis with simpler models and gradually add more complex components. As done for the \textit{NuSTAR} data alone, we start by applying a simple absorbed powerlaw model, which gives a relatively poor fit ($\chi^2/\nu$\,=\,$\frac{986}{888}$\,=\,1.11) with large residuals present mainly at lower energies. Values for the interstellar absorption and for the powerlaw index remained within the same range, i.e., $N_H$ ($\times$\,10$^{22}$)\,=\,12.8\,$\pm$\,0.1\,cm$^{-2}$ and $\Gamma$ =  1.75\,$\pm$\,0.01. Cutoff powerlaw models such as \texttt{cutoffpl} or the combination  \texttt{highecut*powerlaw} provide similar results and set lower limits for the cutoff energy of $\sim$\,230\,keV and $\sim$\,216\,keV, respectively. A broken powerlaw (\texttt{bknpo}) improves the fit ($\chi^2/\nu$\,=\,$\frac{939}{886}$\,=\,1.06) and gives indices of $\Gamma$ = 1.51\,$\pm$\,0.10 and 1.75\,$\pm$\,0.02 with a break energy of 5.4\,$\pm$\,0.5\,keV, suggesting the need of another component at lower energies. 

A significant fit improvement ($\chi^2/\nu$\,=\,$\frac{923}{886}$\,=\,1.04) is then achieved when a multicolour blackbody disc is included in the model (F-test probability of $\sim$\,10$^{-13}$). The \texttt{diskbb} model has only two parameters: the inner disc temperature energy and a normalisation factor. The former is found to be 0.19\,$\pm$\,0.05\,keV for our fit, value in agreement with previously reported values (see, e.g., \citealp{cas141}). From the latter parameter -- the normalisation factor -- the inner disc radius may be estimated, as its value is defined by $norm$ $\sim$\,(R$_{\text{in}}$/D$_{10\,\text{kpc}}$)$^2$\,cos\,$\theta$. Adopting the distance to the Galactic Centre -- 8.5\,kpc -- as the distance to \E\space and an inclination of 60$^{\circ}$, the lower and upper limits for the normalisation reveal inner radii varying roughly from 25\,R$_g$ to 40\,R$_g$ for a 10\,M$_{\odot}$ mass black hole. These limits are in agreement with the radii values reported by \cite{2010AIPC.1248..189R} and \cite{2014ApJ...780...63N} and support the growing body of evidence contradicting the standard idea in which the disc is truncated at \textit{very} large radii for sources in the LHS. From this model we have also calculated an unabsorbed flux (2--200\,keV) of $\approx$\,3.2\,$\times$\,10$^{-9}$\,erg\,$\cdot$\,cm$^{-2}$\,$\cdot$\,s$^{-1}$, which -- for the same distance and mass previously assumed -- corresponds to $\sim$\,2\% of the Eddington luminosity.

\subsection{Comptonisation Models}

Maintaining the \texttt{diskbb} component we tried a few comptonisation  models \footnote{For all models with the seed photons temperature as a parameter, we have assigned it to that of the disc temperature from \texttt{diskbb}, i.e., $kT_{\text{seed}}\, \text{=}\, kT_{\text{diskbb}}$.} for the hard part of the spectrum. We start with \texttt{comptt} \citep{1994ApJ...434..570T}, a model describing comptonisation of blackbody-like seed photons in a hot plasma. Unlike reported by \cite{2009ApJ...693.1871B} and \cite{cas141}, this model provides a very high plasma temperature $kT_e$\,>\,276\,keV and a very low optical depth $\tau$ of 0.08$\pm$\,0.02. Although these values are consistent with the $\tau-kT_e$ degeneracy characteristic for a thermal comptonisation spectrum (see, e.g., \citealp{2008MmSAI..79..118P}), they are not typical for black holes in the LHS. None the less -- and in detriment of fit quality ($\chi^2/\nu$\,=\,$\frac{924}{885}$\,=\,1.04 against $\chi^2/\nu$\,=\,$\frac{949}{886}$\,=\,1.07) -- by freezing parameter $\tau$ at 1 we can retrieve the expected plasma temperature of $\sim$\,50\,keV (i.e., as reported by studies previously mentioned). The \texttt{nthcomp} model \citep{1996MNRAS.283..193Z} provides no noticeable difference in the fit quality ($\chi^2/\nu$\,=\,$\frac{923}{885}$\,=\,1.04) and the electron temperature also pegs at the model hard limit, however with a lower limit value ($kT_e$\,>\,115\,keV).

We also apply \texttt{compps} \citep{1996ApJ...470..249P}, which computes comptonisation spectra for different geometries and takes into account reflection from the cold disc. By assuming a spherical and Maxwellian electron distribution for the corona (i.e., only thermal emission), a non-ionised disc ($\xi$ = 0), a relative iron and metal abundances varying only between 1 and 3 (based on previous reports by \citealp{1999ApJ...520..316S} and \citealp{2010AIPC.1248..189R}) and a viewing angle of 60$^{\circ}$, the model sets constraints on the plasma temperature: 222\,keV\,$\leq$\,$kT_e$\,$\leq$\,426\,keV with an optical depth of 0.19\,$\leq$\,$\tau$\,$\leq$\,0.48. A reflection component up to $\Omega/2\pi$\,$\sim$\,0.38 is also detected and will be discussed in the following subsection.
This latter model provides the best fit so far ($\chi^2/\nu$\,=\,$\frac{911}{882}$\,=\,1.03) -- which is further improved ($\chi^2/\nu$\,=\,$\frac{901}{882}$\,=\,1.02) if a partially ionised disc ($\xi$ = 1000\,erg\,$\cdot$\,cm\,$\cdot$\,s$^{-1}$) is considered, yielding values closer to those expected for the source state: 48\,keV\,$\leq$\,$kT_e$\,$\leq$\,282\,keV and 0.67\,$\leq$\,$\tau$\,$\leq$\,2.89 (also an amount of reflection $\Omega/2\pi$\,$\lesssim$\,0.19).

Additionally, we apply the \texttt{bmc} model \citep{1997ApJ...487..834T}, which considers a more general case where the comptonisation spectrum may be either from a hot corona or from bulk-motion upscattering. Since this model is described as a self-consistent convolution (rather than an additive combination) of powerlaw and seed photons spectrum, the \texttt{diskbb} component may be removed. The output parameters for the \texttt{bmc} fit ($\chi^2/\nu$\,=\,$\frac{923}{886}$\,=\,1.04)  are consistent with those found for the \texttt{diskbb+powerlaw} model, i.e., $kT_{\text{seed}}$\,$\approx$\,0.17\,keV and $\alpha$\,$\approx$\,0.76, where $\alpha$ = $\Gamma$\,$-$\,1. The model has another parameter, $A$, that is related to the ratio of photons upscattered via bulk-motion to the photons thermally comptonised in the corona (see, e.g., \citealp{1999ApJ...517..367B}). The limits provided for this parameter indicate practically none (< 0.5\%) to 20\% dynamically upscattered photons, i.e., this mechanism is not the dominant for our spectrum. 
 
\begin{table*}
\caption{Spectral Fit Parameters}
\label{tab:02}
\resizebox{\textwidth}{!}{
\hspace*{-0.65cm}
\begin{tabular}{lccccccccccc}
\hline
\multirow{2}{*}{Model/Parameter}            & N$_\text{H}$                        & \multirow{2}{*}{$\Gamma$} & T$_{\text{in}}$ & \multirow{2}{*}{$\Gamma_2$}   & E$_{\text{fold/break}}$       & kT$_{\text{e}}$           & \multirow{2}{*}{$\tau$}       & \multirow{2}{*}{log$_A$}       & $\xi$                                              & Reflection                    & \multirow{2}{*}{$\chi^2/\nu/\chi^2_r$} \\
                                  & ($\times\,10^{22}\,\text{cm}^{-2}$) &                           & (keV)           &                               & (keV)                         & (keV)                     &                               &                                & ($\text{erg}\,\cdot\text{cm}\,\cdot\text{s}^{-1}$) & ($\Omega/2\pi$)               &                               \\ \hline
\texttt{powerlaw}                 & 12.8$\pm\,0.1$                      & 1.75$\pm\,0.01$           & -               & -                             & -                             & -                         & -                             & -                              & -                                                  & -                             & 986/888/1.11                     \\
\texttt{cutoffpl}                 & 12.7$\pm\,0.1$                      & 1.72$\pm\,0.02$           & -               & -                             & $>$\,233                      & -                         & -                             & -                              & -                                                  & -                             & 978/887/1.10                       \\
\texttt{highecut*powerlaw}        & 12.7$\pm\,0.1$                      & 1.71$\pm\,0.02$           & -               & -                             & $>$\,215                      & -                         & -                             & -                              & -                                                  & -                             & 973/886/1.10                       \\
\texttt{bknpo}                    & 11.9$\pm\,0.3$                      & 1.51$\pm\,0.01$           & -               & 1.75$\pm\,0.02$               & 5.43$^{\text{+}0.71}_{-0.20}$ & -                         & -                             & -                              & -                                                  & -                             & 939/886/1.06                       \\
\texttt{diskbb+powerlaw}          & 13.3$\pm\,0.2$                      & 1.77$\pm\,0.01$           & 0.19$\pm\,0.05$ & -                             & -                             & -                         & -                             & -                              & -                                                  & -                             & 923/886/1.04                       \\
\texttt{diskbb+comptt}            & 13.2$\pm\,0.2$                      & -                         & 0.18$\pm\,0.05$ & -                             & -                             & $>$\,276                  & $<$\,0.1                      & -                              & -                                                  & -                             & 924/885/1.04                       \\
\texttt{diskbb+comptt}            & 12.8$\pm\,0.1$                      & -                         & 0.15$\pm\,0.05$ & -                             & -                             & 46.7$\pm\,1.2$            & 1*                            & -                              & -                                                  & -                             & 949/886/1.07                       \\
\texttt{diskbb+nthcomp}           & 13.3$\pm\,0.2$                      & 1.77$\pm\,0.01$           & 0.19$\pm\,0.05$ & -                             & -                             & $>$\,114                  & -                             & -                              & -                                                  & -                             & 923/885/1.04                       \\
\texttt{diskbb+compps}            & 12.4$\pm\,0.2$                      & -                         & 0.15$\pm\,0.05$ & -                             & -                             & 325$^{\text{+}146}_{-63}$ & 0.30$^{\text{+}0.22}_{-0.13}$ & -                              & 0*                                                 & 0.27$^{\text{+}0.21}_{-0.06}$ & 911/882/1.03                       \\
\texttt{diskbb+compps}            & 13.2$\pm\,0.2$                      & -                         & 0.17$\pm\,0.05$ & -                             & -                             & 82$^{\text{+}200}_{-34}$  & 1.69$^{\text{+}1.2}_{-0.9}$   & -                              & 1000*                                              & 0.13$^{\text{+}0.05}_{-0.03}$ & 901/882/1.02                       \\
\texttt{bmc}                      & 13.3$\pm\,0.2$                      & $^a$1.77$\pm\,0.01$\phantom{$^a$}           & 0.17$\pm\,0.05$ & -                             & -                             & -                         & -                             & -1.47$^{\text{+}0.80}_{-1.00}$ & -                                                  & -                             & 923/886/1.04                       \\
\texttt{diskbb+pexrav}            & 13.2$\pm\,0.3$                      & 1.75$\pm\,0.03$           & 0.19$\pm\,0.05$ & -                             & -                             & -                         & -                             & -                              & $^b$-\phantom{$^a$}                                                  & 0.12$^{\text{+}0.08}_{-0.06}$ & 920/882/1.04                       \\
\texttt{diskbb+pexriv}            & 13.1$\pm\,0.2$                      & 1.75$\pm\,0.02$           & 0.18$\pm\,0.05$ & -                             & -                             & -                         & -                             & -                              & 1000*                                              & 0.16$^{\text{+}0.06}_{-0.03}$ & 904/882/1.02                       \\
\texttt{diskbb+pexriv}            & 12.9$\pm\,0.2$                      & 1.74$\pm\,0.02$           & 0.16$\pm\,0.05$ & -                             & -                             & -                         & -                             & -                              & 5000$^{\text{+}0}_{-3592}$                         & 0.12$^{\text{+}0.09}_{-0.03}$ & 899/881/1.02                       \\
\texttt{diskbb+powerlaw+reflionx} & 13.0$\pm\,0.2$                      & 1.71$\pm\,0.05$           & 0.16$\pm\,0.05$ & -                             & -                             & -                         & -                             & -                              & 6480$^{\text{+}0}_{-3703}$                         & -                             & 904/883/1.02                       \\
\texttt{diskbb+xillver}           & 12.9$\pm\,0.1$                      & 1.70$\pm\,0.02$           & 0.16$\pm\,0.05$ & -                             & -                             & -                         & -                             & -                              & $^c$4897$^{\text{+}2515}_{-1878}$\phantom{$^a$}       & 0.25$^{\text{+}0.14}_{-0.06}$ & 904/882/1.02                       \\
\texttt{diskbb+xillver+comptt}    & 13.0$\pm\,0.2$                      & -                         & 0.16$\pm\,0.05$ & $^d$1.85$^{\text{+}0.22}_{-0.14}$\phantom{$^a$} & -                             & 327$^{\text{+}163}_{-54}$ & 0.10$^{\text{+}1.58}_{-0.01}$ & -                              & $^e$5248$^{\text{+}2880}_{-2493}$\phantom{$^a$}       & -1*                             & 900/881/1.02                       \\  \hline
\end{tabular}}
{\begin{flushleft}\small{\bf{Notes:}}  Errors are at 90\% confidence limit determined via the \texttt{error} command in \textsc{xspec}. All models include a multiplicative constant and the interstellar absorption model \texttt{phabs}. For models in which the relative iron/metal abundances are free parameters, we set them to vary  between 1--3, as well as the system's inclination was fixed at 60$^{\circ}$ (see text for details). An asterisk (*) next to a parameter means that it was frozen at that value. Errors +0 indicate that the parameter reached the allowed hard limit for the model. 

$^a$The spectral index of \texttt{bmc} is actually parametrised by $\alpha$ (=\,0.77) and $\Gamma\,\text{=} \,\alpha\,\text{+}\,1$. 

$^b$\texttt{pexrav} describes reflection from neutral material, i.e., equivalent to \texttt{pexriv} for $\xi$\,=\,0. 

$^{c,e}$The output ionisation parameter is provided by \texttt{xillver} as log\,$\xi$; the original output values were 3.69$^{\text{+}0.18}_{-0.21}$ and 3.72$^{\text{+}0.19}_{-0.28}$, respectively. 

$^d$As for this combination the \texttt{xillver} model is set to return only the reflected component, this $\Gamma$ refers to the reflected photon index. \end{flushleft}}
\end{table*}

\subsection{Non-Relativistic Reflection Models}

The fact that the best fit from the comptonisation models occurs for \texttt{compps} -- a model that includes a reflection component -- suggests that this component may be present in the data and needs to be taken into account.
Model \texttt{compps} follows the computational method of the widely used angle-dependent non-relavistic reflection models \texttt{pexrav} and \texttt{pexriv} \citep{1995MNRAS.273..837M}. 
By applying these two models, for the same initial conditions we set for \texttt{compps},  we find reflection fraction values around 20\% for either neutral (\texttt{pexrav}) or partially ionised (\texttt{pexriv}, $\xi$ = 1000\,erg\,$\cdot$\,cm\,$\cdot$\,s$^{-1}$) discs. As it happened for \texttt{compps}, the scenario with a partially ionised disc provides a better fit ($\chi^2/\nu$\,=\,$\frac{904}{882}$\,=\,1.02 against $\chi^2/\nu$\,=\,$\frac{920}{882}$\,=\,1.04). If ionisation parameter is let free, best fit from \texttt{pexriv} sets it to its hard limit, 5000\,erg\,$\cdot$\,cm\,$\cdot$\,s$^{-1}$.

When applying the constant density ionised disc model \texttt{reflionx} \citep{2005MNRAS.358..211R} -- which does not provide a reflection fraction -- an ionisation of $\approx$\,6500\,erg\,$\cdot$\,cm\,$\cdot$\,s$^{-1}$ is obtained. As \texttt{reflionx} is a pure reflection model, a \texttt{powerlaw} was included to represent the continuum and their photon indices tied. The reflection code \texttt{xillver} \citep{2013ApJ...768..146G} yields a reflection fraction $R\,\approx\,$\,0.25 and an ionisation parameter of $\approx\,$\,4900 (log\,$\xi$\,=\,3.69). This latter uses the coronal geometry and considers the incident spectrum to be a cutoff powerlaw. A different flavor of the model, \texttt{xillvercp}, allows for a \texttt{nthcomp} incident spectrum; however, the soft seed photons energy is not a free parameter and is set to be $kT_{\text{seed}}$ = 0.05\,keV. Thus, instead, we add the \texttt{comptt} model to describe the incident comptonisation continuum and set \texttt{xillver} to return only the reflected component (by fixing the reflection fraction to $-$1). In this case, we tie the energy cutoff of the reflected spectrum to twice the $kT$ temperature from \texttt{comptt}. The output comptonisation values for this combination are consistent with those found for the \texttt{comptt} alone and the fit quality slightly improved (F-test probability of $\sim$\,10$^{-5}$). The results for the models applied up to now are presented in Table \ref{tab:02}.

\subsection{Relativistic Reflection Models}

\begin{figure}
	\includegraphics[width=\columnwidth]{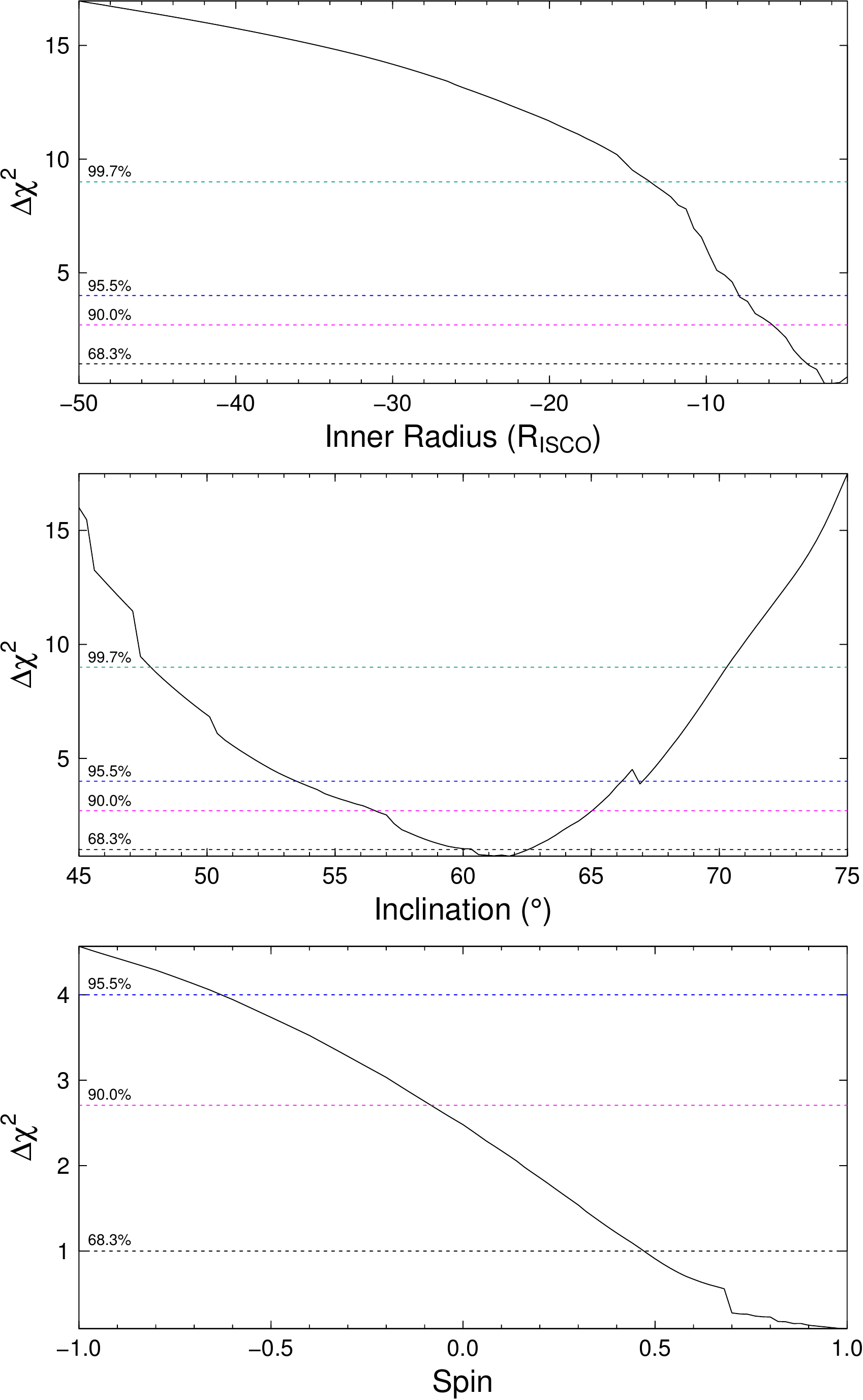}
    \caption{$\Delta\chi^2$ vs parameter value for the inner radius (upper panel), inclination (middle panel) and spin (bottom panel) for the model \texttt{diskbb+powerlaw+relconv(reflionx)}. The dashed lines correspond to the confidence levels at 68.3\% (1\,$\sigma$; black), 90.0\% (1.6\,$\sigma$; magenta), 95.5\% (2\,$\sigma$; blue) and 99.7\% (3\,$\sigma$; bluish green). \textbf{Note:} The negative values in the horizontal axis of the upper panel are merely a model convention and indicate that \texttt{relconv} was set to calculate the radius in R$_{\textsc{isco}}$ units.   
        }
    \label{fig:05}
\end{figure}

To explore the parameter space for spin and inclination we test a number of relativistic reflection models, which are the essence of the X-ray reflection spectroscopy method \textit{per se}. We start with our previously mentioned fit of \texttt{reflionx}, now convolved with \texttt{relconv} \citep{2010MNRAS.409.1534D} to allow for relativistic effects (\texttt{diskbb+powerlaw+relconv(reflionx)} in \textsc{xspec} notation). As before, the photon index of \texttt{reflionx} is tied to that of the \texttt{powerlaw} model. For \texttt{relconv}, we fix the disc outer radius at $R_{\text{out}}$\,=\,400\,R$_g$ and assume the coronal emissivity profile $\epsilon$ = $r^{-q}$  to be unbroken with an emissivity index $q$ = 3 (i.e., $q_{\text{in}}$ = $q_{\text{out}}$ = 3). The disc inner radius is initially let frozen at -1, meaning it extends all the way to the R$_{\textsc{isco}}$ (radius of the innermost stable circular orbit). The spin and inclination are let free to vary from their model initial default values, i.e., 0.998 and 30$^{\circ}$, respectively. Although the best fit ($\chi^2/\nu$\,=\,$\frac{897}{881}$\,=\,1.02) for this configuration leaves the spin parameter at its maximum value, $a_*$ = 0.998, the parameter reaches both its hard limits (-0.998 and 0.998) when we attempt to calculate the 90\% confidence limits. The inclination, however, rapidly increases and is well defined at 63.7$^{\text{+}4.6}_{-7.9}$\,degrees. To investigate how these two parameters behave for different inner radii, we perform further fits for the disc radius fixed at distances (in R$_{\textsc{isco}}$ units)  of 10, 20, 50  and also free to vary. As the radius increases, the quality of the fit is marginally worsened ($\chi^2$ = 899, 903 and 909, respectively) and the inclination becomes higher and somewhat looser (75.2$^{\text{+}10.6}_{-18.9}$, 77.0$^{\text{+}10.4}_{-22.9}$  and 79.1$^{\text{+}9.5}_{-52.3}$ degrees). The parameter spin remained at $a_*$\,$\simeq$\,0.998 for all three scenarios but failed again in providing 90\%  level limits. For the case where $R_{\text{in}}$ is free, the best fit ($\chi^2/\nu$\,=\,$\frac{896}{880}$\,=\,1.02) sets the inner radius to $\sim$\,2.5 ($\lesssim$\,6\,R$_{\textsc{isco}}$  at 90\% confidence)  and provides an inclination of 61.2$^{\text{+}3.9}_{-4.9}$\,($^{\circ}$);  the spin reaches its maximum value but is again not well constrained. Figure \ref{fig:05} shows the $\Delta\chi^2$ variation for these three parameters, calculated via the \texttt{steppar} command from \textsc{xspec}. Confidence levels at 68.3\% (1\,$\sigma$; black dashed line), 90.0\% (1.6\,$\sigma$; magenta dashed line), 95.5\% (2\,$\sigma$; blue dashed line) and 99.7\% (3\,$\sigma$; bluish green dashed line) are indicated. According to the plots, an inner radius truncated at over $\sim$\,15\,R$_{\textsc{isco}}$ (upper panel) and a disc inclination below $\sim$\,45$^{\circ}$ or above $\sim$\,70$^{\circ}$ (middle panel) may be ruled out at 3\,$\sigma$ confidence level for this model. The spin parameter (bottom panel) clearly favours the hard limit (> 0.5 at 1\,$\sigma$) but still admits any value within 3\,$\sigma$. 
To test whether the disc and the observer see different continua -- as pointed out, e.g., by \cite{2015ApJ...808..122F} and \cite{2016ApJ...821L...6P} for GX\,339--4 -- we repeat the last fit (with $R_{\text{in}}$ free to vary) leaving the photon indices of the \texttt{powerlaw} and \texttt{reflionx} models untied. The powerlaw as seen from the reflector in fact assumes a slightly different index but is also more relaxed and yet in the same range ($\Gamma\sim$\,2; 1.5 $\lesssim\,\Gamma\,\lesssim\,$2.5). Contrary to reported in the mentioned studies of GX\,339--4 (where a good fit is only achieved for different powerlaw indices), this additional free parameter does not improve our fit ($\chi^2/\nu$\,=\,$\frac{896}{879}$\,=\,1.02) nor causes significant changes in the quoted results.

\begin{figure}
	\includegraphics[width=\columnwidth]{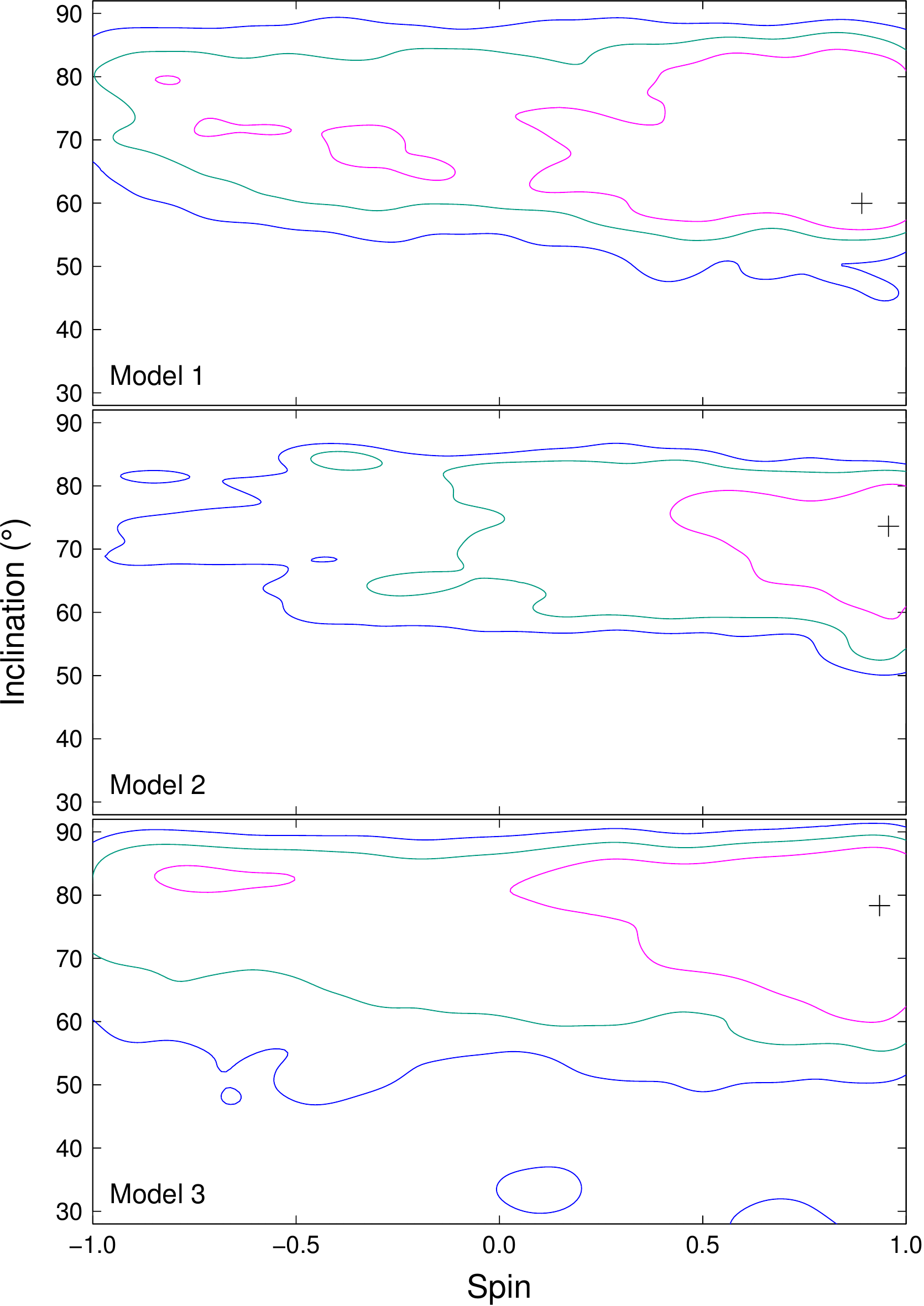}
    \caption{MCMC output distribution for parameters spin and inclination for combinations \texttt{relxill+comptt} (Model 1; upper panel), \texttt{relxilllp+comptt} (Model 2; middle panel) and \texttt{relxilllp} alone (Model 3; bottom panel). The black plus symbols indicate the best-fitting values. Confidence contours  of 1 (magenta), 2 (bluish green) and 3 (blue) $\sigma$ are shown.}
    \label{fig:06}
\end{figure}

\begin{figure}
	\includegraphics[width=\columnwidth]{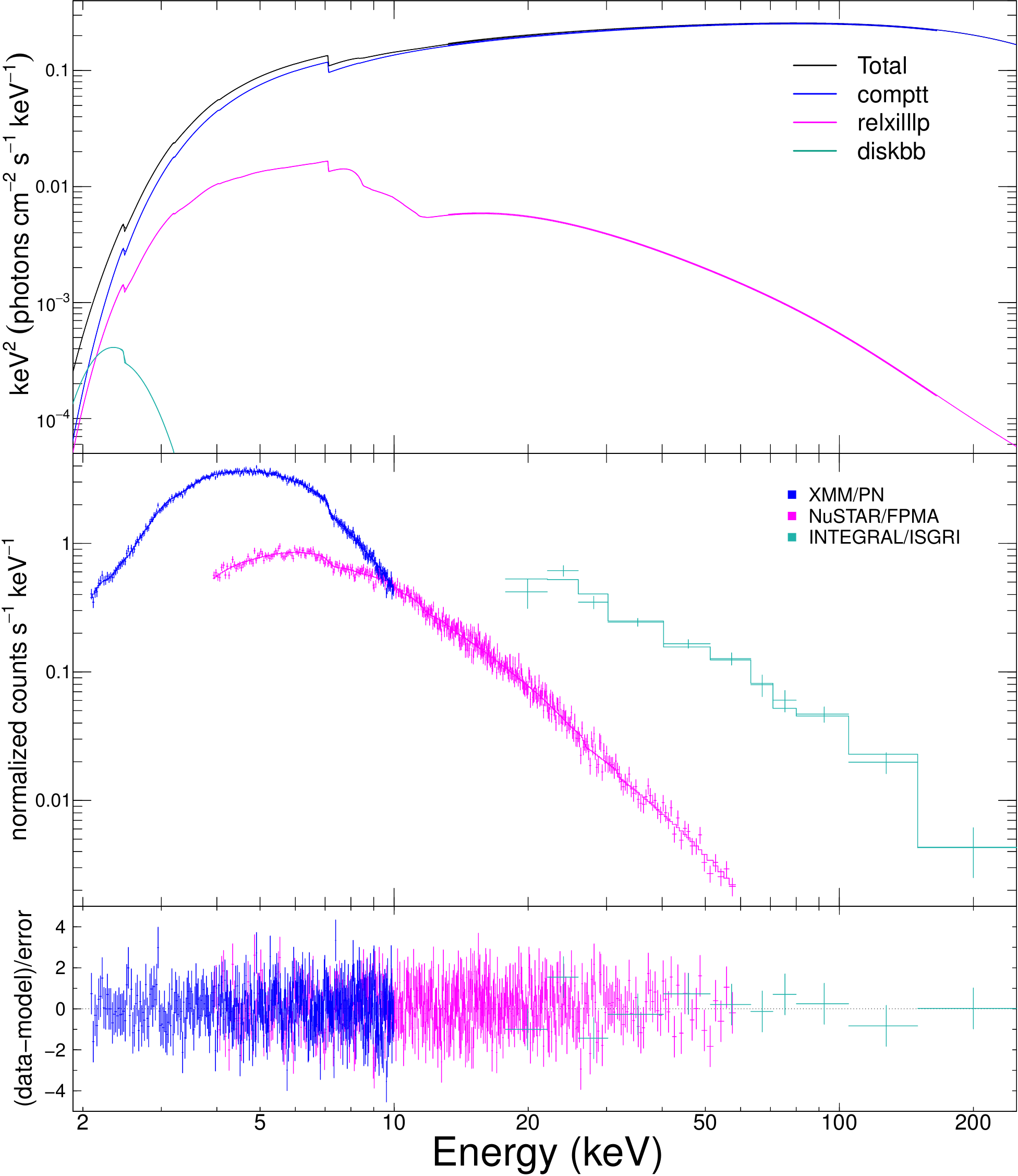}
    \caption{Model components (upper panel), model fitted to the data (middle panel)  and its residuals (bottom panel) for Model 2. \textbf{Note:} Vertical axis of bottom panel is in \textit{(data-model)/error}, equivalent to $\Delta\,\chi$, so magnitudes of \textit{XMM} (blue), \textit{NuSTAR} (magenta) and \textit{INTEGRAL} (bluish green) residuals are comparable.}
    \label{fig:07}
\end{figure}

\begin{figure*}
	\includegraphics[width=\textwidth]{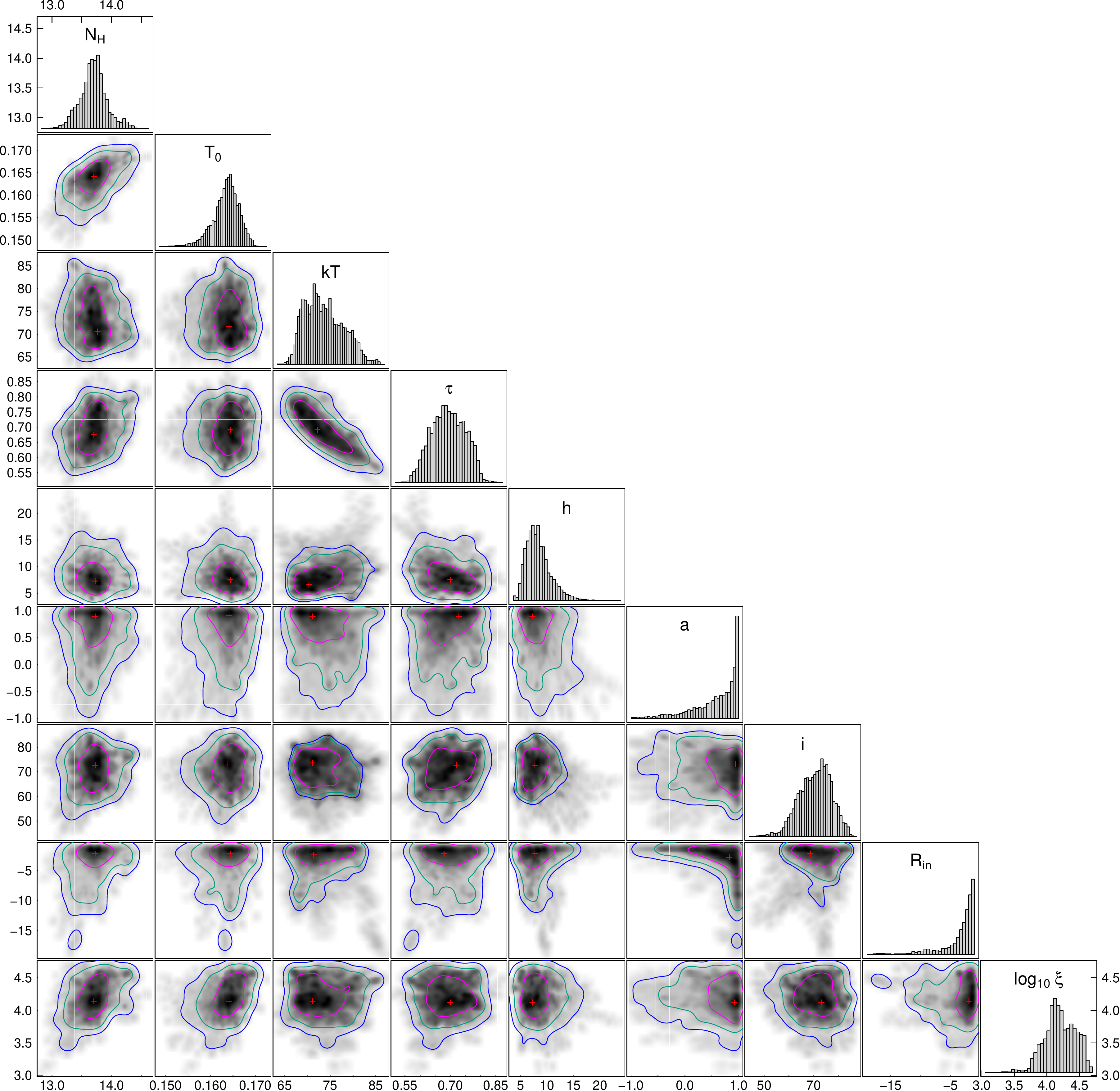}
    \caption{Joint distribution of selected parameters for Model 2, calculated from the MCMC output. The red plus  symbols indicate the best-fitting values. Confidence contours  of 1 (magenta), 2 (bluish green) and 3 (blue) $\sigma$ are shown. \textbf{Note:} Parameters and their units are: $N_H$ ($\times$\,10$^{22}$\,cm$^{-2}$), $T_0$ (keV), $kT_e$ (keV), $h$ (R$_{\text{g}}$), $i$ (degrees), $R_{in}$ (R$_{\textsc{isco}}$), $\text{log}\,\xi$ (log erg\,$\cdot$\,cm\,$\cdot$\,s$^{-1}$). $\tau$ and $a$ are dimensionless. }
    \label{fig:08}
\end{figure*}

To proceed with the spin and inclination analysis we apply two different coronal geometries available from the \texttt{relxill} model family (\citealp{2014MNRAS.444L.100D}; \citealp{2014ApJ...782...76G}), i.e.,  the standard coronal geometry (\texttt{relxill} model itself) and the lamp post geometry (\texttt{relxilllp}). For the following results, the associated errors are calculated by a Markov Chain Monte Carlo (MCMC) algorithm from \cite{2013PASP..125..306F}, implemented for \textsc{xspec} by Jeremy Sanders\footnote{https://github.com/jeremysanders/xspec\_emcee}.
As previously done for the non-relavistic version \texttt{xillver}, we use \texttt{comptt} to describe the primary continuum. For consistency with \texttt{relconv(reflionx)}, which also assumes the standard geometry, we adopt the same conditions for \texttt{relxill} ($q_{\text{in}}$ = $q_{\text{out}}$ = 3 and $R_{\text{out}}$\,=\,400\,R$_g$). For \texttt{relxilllp}, that assumes the corona to be a point source at a certain height $h$ above the black hole, we leave this parameter free from its default value (6\,R$_g$). We again perform fits for inner radii (in R$_{\textsc{isco}}$ units)  at 1, 10, 20, 50 and free. The overall outcome with increasing radius, for both combinations, is somewhat similar: the quality of the fit decreases, the disc inclination tends to assume higher values and the spin -- still not well constrained -- approaches the upper limit for all cases. These variations are presented in Table \ref{tab:03} for the two models (Model 1: \texttt{diskbb+relxill+comptt}; Model 2: \texttt{diskbb+relxilllp+comptt}). 
\begin{table}
\caption{Inclination and fit quality for different radii}
\label{tab:03}
\resizebox{\columnwidth}{!}{
\hspace*{-0.65cm}
\begin{tabular}{clcclll}
\cline{1-6}
\multirow{2}{*}{Model}  & \multicolumn{1}{c}{\multirow{2}{*}{Parameter}} & \multicolumn{4}{c}{R$_{\text{in}}$ (R$_{\textsc{isco}}$)}                                                                     &  \\
                        & \multicolumn{1}{c}{}                           & 1                            & 10                            & \multicolumn{1}{c}{20}        & \multicolumn{1}{c}{50}        &  \\ \cline{1-6}
\multirow{2}{*}{\bf{1}} & Inclination ($^{\circ}$)                       & 60.8$^{\text{+}18.3}_{-5.7}$ & 72.5$^{\text{+}11.3}_{-12.8}$ & 77.8$^{\text{+}9.1}_{-25.4}$ & 84.9$^{\text{+}1.5}_{-63.8}$  &  \\
                        & $\chi^2/\nu$                                   & 897/879                      & 899/879                       & 906/879                       & 915/879                       &  \\ \cline{1-6}
\multirow{2}{*}{\bf{2}} & Inclination ($^{\circ}$)                       & 66.2$^{\text{+}8.2}_{-10.2}$ & 71.3$^{\text{+}9.4}_{-12.5}$ & 72.7$^{\text{+}6.6}_{-8.6}$ & 78.4$^{\text{+}4.7}_{-9.7}$ &  \\
                        & $\chi^2/\nu$                                   & 899/878                      & 901/878                       & 906/878                       & 907/878                       &  \\ \cline{1-6}
\end{tabular}}
\end{table}
For many parameters no considerable changes between Models 1 and 2 occurred: the ionisation assumes a relatively broad range of 2.3\,$\lesssim$\,$\text{log}\,\xi$\,$\lesssim$\,3.7 (log erg\,$\cdot$\,cm\,$\cdot$\,s$^{-1}$) for any radius for both models and the iron abundance raises from the lower to the higher limit we imposed (1 to 3) as the radius increases. For \texttt{relxilllp}, the source height varies roughly within the range of 4\,R$_g$\,$\lesssim$\,$h$\,$\lesssim$\,13\,R$_g$ -- including errors at 90\% confidence level -- with the higher values happening for larger radii. The powerlaw indices remain in the same range as that of \texttt{reflionx} when left free (1.5 $\lesssim\,\Gamma\,\lesssim\,$2.5), with no apparent trend with increasing radius for the coronal geometry and with the powerlaw becoming harder (approaching the ``observer index'') for farther inner radii in the lamp-post geometry. It is interesting to mention that this latter geometry places the comptonisation parameters in the expected range for \E, i.e., 30\,keV\,$\lesssim$\,$kT_e$\,$\lesssim$\,80\,keV and 0.5\,$\lesssim$\,$\tau$\,$\lesssim$\,2.2 for any radius.  
For $R_{\text{in}}$ free, the best fit for Model 1 ($\chi^2/\nu$\,=\,$\frac{896}{878}$\,=\,1.02) and Model 2 ($\chi^2/\nu$\,=\,$\frac{897}{877}$\,=\,1.02) yield radii of 2.1$^{\text{+}4.2}_{-0.3}$ and 5.2$^{\text{+}4.0}_{-3.1}$ R$_{\textsc{isco}}$, respectively.  We have also performed another fit letting \texttt{relxilllp} represent both incident and reflected spectra alone (Model 3: \texttt{diskbb+relxilllp}), in such a way that we can leave the reflection fraction free. The best fit ($\chi^2/\nu$\,=\,$\frac{897}{879}$\,=\,1.02), which sets the inner radius at 1.7$^{\text{+}9.2}_{-0.6}$ R$_{\textsc{isco}}$, provides a reflection fraction of $R$ = 0.26$^{\text{+}0.27}_{-0.15}$. The powerlaw index -- now unique for both incident and reflected spectrum -- is $\Gamma$ = 1.78$^{\text{+}0.02}_{-0.05}$. Other parameters, which are common to Models 1 and 2, have not changed much from the values quoted previously. In regard of spin and inclination, Figure \ref{fig:06} presents the MCMC output distribution for these two parameters for the cases where $R_{\text{in}}$ was left free in Models 1, 2 and 3. For all three models the spin behaviour differs very little from that of \texttt{relconv(reflionx)}, displayed in the bottom panel of Figure \ref{fig:05}: it reaches the hard limit but is only poorly suggestive it might be $\gtrsim$\,0.5 (1\,$\sigma$); any value is still possible within 3\,$\sigma$. As for the inclinations, they all coincide to be in the range of $\sim$\,60--85$^{\circ}$ (2\,$\sigma$); values below $\sim$\,55$^{\circ}$ can be excluded with 3\,$\sigma$ confidence for Models 1 and 2, whereas for Model 3 this same confidence region encompass a broader range, with marginally lower inclinations. Model components, model fitted to the data and its residuals for Model 2 are displayed in Figure \ref{fig:07}. Aside from providing the best constraints on spin and inclination, the reason we elect to show this combination is that it includes the comptonisation component, in which best-fitting values of $\tau$ and $kT_e$ are -- as formerly discussed -- in a more likely interval for \E. For a comprehensive description of the model variables and dependencies, we also present, in Figure \ref{fig:08}, the joint distributions of selected parameters for Model 2.   

\subsection{A Putative Mass Estimate Effort}

\begin{figure}
	\includegraphics[width=\columnwidth]{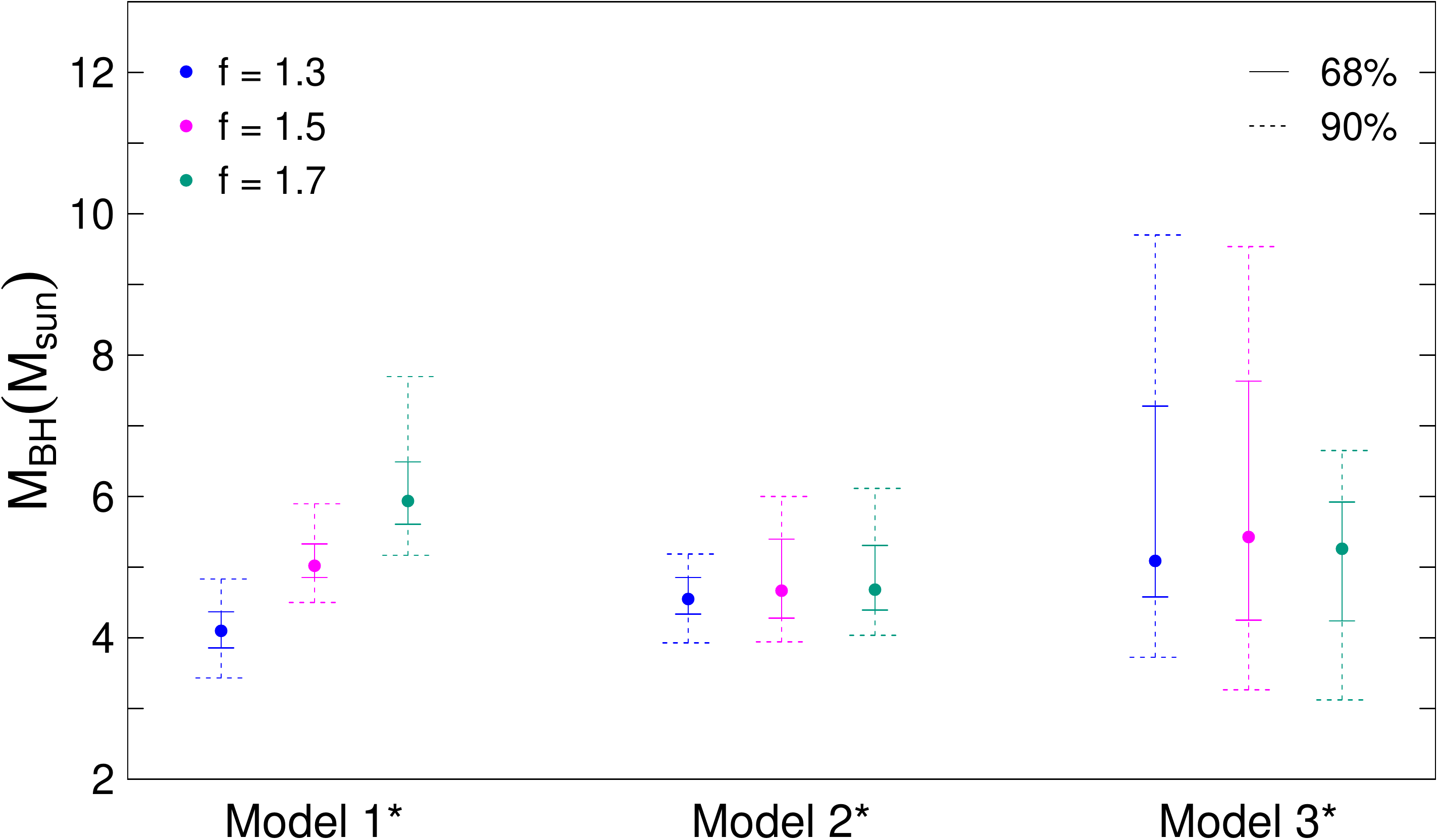}
    \caption{Output mass values for Models 1*, 2* and 3* for spectral hardening factors of $f$ = 1.3 (blue), 1.5 (magenta) and 1.7 (bluish green). Solid and dashed bars indicate 68\% and 90\% confidence respectively (see text for details).}
    \label{fig:09}
\end{figure}

In an attempt to estimate the mass of the black hole, we replace the component characterising the disc spectrum for these three last models (with parameter $R_{\text{in}}$ free) from \texttt{diskbb} to \texttt{kerrbb} \citep{2005ApJS..157..335L}. The latter describes a multitemperature blackbody accretion disc around a Kerr black hole and represents the continuum-fitting method, which consists of the analysis of the thermal spectrum of a geometrically thin and optically thick disc in order to measure the black hole spin (see, e.g., \citealp{2018AnP...53000430B}). Not only this method relies on an inner disc radius close to ISCO, its application depends on prior independent measurements of many parameters -- usually inferred from investigations in other wavelengths -- such as the distance to the source, the inclination of the disc, the accretion rate and the mass of the black hole. However, as we are interested in the black hole mass, we follow the approach performed by \cite{2016ApJ...821L...6P} for GX\,339--4 and use the inclination and spin values provided by the reflection model \textit{as input} parameters for \texttt{kerrbb}. Namely, we tie the spin and inclination of \texttt{kerrbb} to those from the \texttt{relxill} models. Apropos of other parameters, we set the distance to the source to be fixed at 8.5\,kpc and the mass accretion rate is left free to vary from an initial value of 10$^{17}$ g\,$\cdot$\,s$^{-1}$, which is roughly $\sim$\,2\% of the Eddington accretion rate for a 10\,M$_{\odot}$ black hole with efficiency $\eta$ = 0.1. Another noteworthy parameter is the spectral hardening or colour correction factor $f$, whose default value -- $f$ = 1.7 -- is usually a good approximation for sources with luminosities of few times 0.1\,L$_{\text{Edd}}$ \citep{1995ApJ...445..780S}. Since for our calculated luminosity of $\sim$\,0.02\,L$_{\text{Edd}}$ this factor most likely assumes lower values (see, e.g, Table 1 from \citealp{2005ApJ...621..372D}), we perform fits with $f$ fixed at 1.3, 1.5 and 1.7. For the model remaining parameters, we adhere to the default scenarios: a disc with zero torque at the inner boundary ($eta$ = 0), effects of self-irradiation are considered ($rflag$ = 1) and the disc emission is assumed to be isotropic ($lflag$ = 0). Figure \ref{fig:09} shows the resulting black hole mass for the three model combinations (Model 1*: \texttt{kerrbb+relxill+comptt}; Model 2*: \texttt{kerrbb+relxilllp+comptt}; Model 3*: \texttt{kerrbb+relxilllp}) for the respective factors $f$. The masses indicated are the median values of the probability density distribution provided by an MCMC run from the best-fitting parameters; 68\% and 90\% confidence level bars shown were calculated from this value. From Figure \ref{fig:09} we can assert that, conservatively, the resulting black hole masses lie within 3 to 10\,M$_\odot$;  Model 2* provides again the best constraints: 3.9\,M$_\odot$\,$\lesssim$\,$M_{\text{BH}}$\,$\lesssim$\,6.1\,M$_\odot$ at 90\% confidence. It can also be noticed that, apart from slight differences in the mass values for Model 1*, no prominent variation occurs for the mass with respect to the spectral hardening. This also extends to the quality of the fit and to the other parameters, both between the same model for different factors and between the models before and after the disc component was replaced. In other words, most (if not all) parameters were insensitive to the spectral hardening and \texttt{kerrbb} model could fit the lower energy part of the spectrum just as well as \texttt{diskbb}. As to the mass accretion rates, values varied roughly from 0.002--0.04 ($\times$\,10$^{17}$ g\,$\cdot$\,s$^{-1}$), regardless of the model or spectral hardening. 

\subsubsection{Fits with the \texttt{simpl} comptonisation model}

\begin{figure}
	\includegraphics[width=\columnwidth]{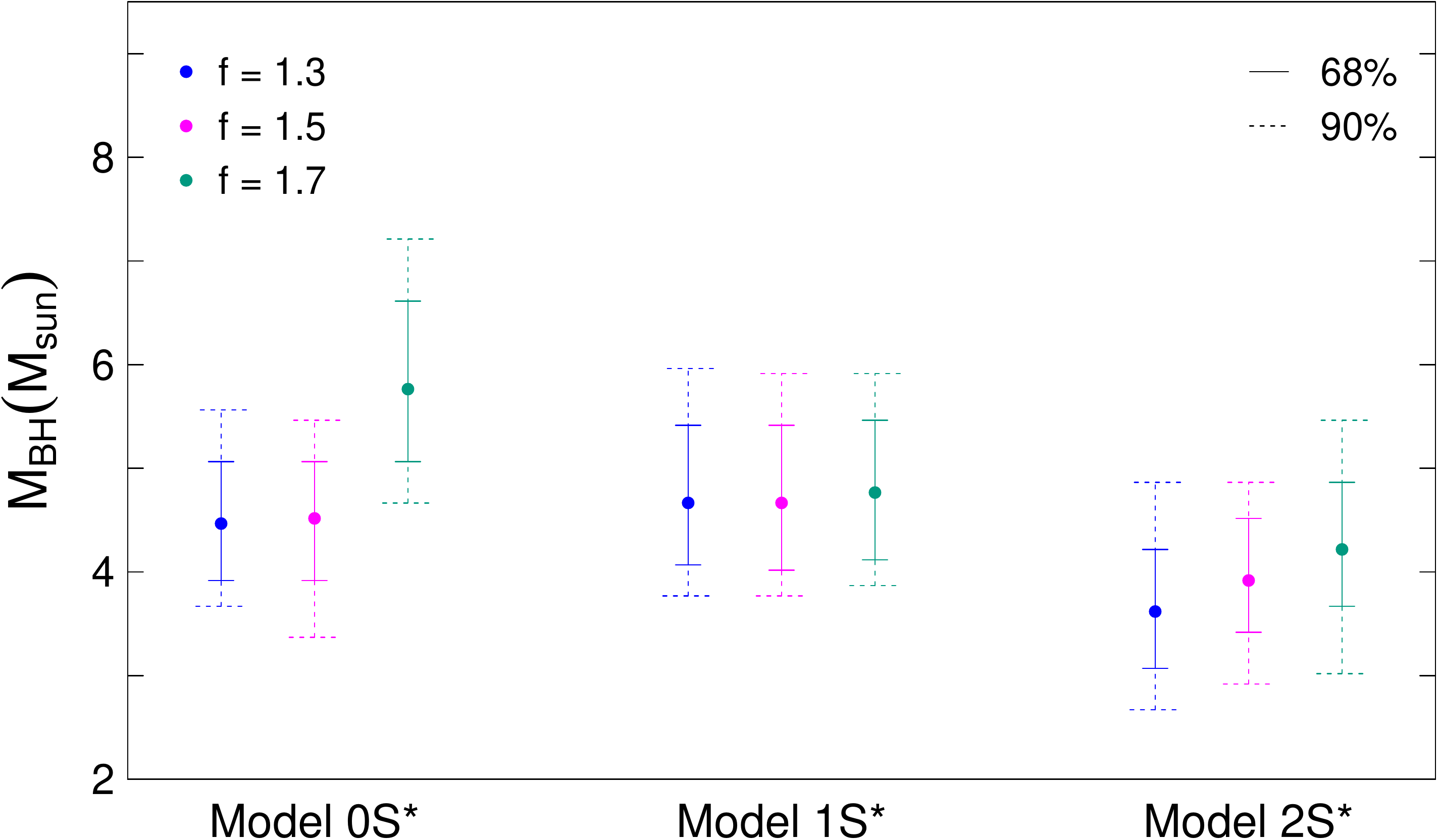}
    \caption{Same as Figure \ref{fig:09}, now for Models 0S*, 1S* and 2S* (see text for details).}
    \label{fig:10}
\end{figure}

It has been pointed out in \cite{2009ApJ...701L..83S} that when applying the continuum-fitting method, the use of additive combinations -- such as a disc plus a powerlaw or plus a comptonisation model -- might be inadequate, as these additive models fail to correctly account for the soft component's contribution to the powerlaw spectrum. Instead, the \texttt{simpl} model \citep{2009PASP..121.1279S}, which self-consistently generates the compton component from the thermal seed photons, should be applied. Therefore, we did the fits of Model 1* and Model 2* again (now to be called Model 1S* and Model 2S*, respectively) replacing the additive combination \texttt{kerrbb+comptt} to the convolution \texttt{simpl(kerrbb)}. We also apply \texttt{simpl(kerrbb)} to the \texttt{relconv(reflionx)} fit, i.e., \texttt{simpl(kerrbb)+relconv(reflionx)} -- which we name Model 0S*. The \texttt{simpl} model is parametrised by only two parameters: the photon index $\Gamma$ and the scattered fraction $f_{\text{SC}}$; the latter, by default, provides the fraction of photons from the accretion disc that are upscattered in energy by the corona. As before, we perform three fits for each, with the spectral hardening $f$ fixed at 1.3, 1.5 and 1.7. We find, again, that no substantial differences are caused by the choice of this parameter. Likewise, the output parameters for the best-fits were all in the same range as the ones quoted in the previous subsections. The resulting masses for all cases are shown in Figure \ref{fig:10} (displayed in the same way of Figure \ref{fig:09}) and some parameters of interest are shown in Table \ref{tab:04}, for the intermediate case of $f$ = 1.5.

\begin{table}
\caption{Best-fit parameters with the \texttt{simpl} model combinations}
\label{tab:04}
\resizebox{\columnwidth}{!}{
\hspace*{-0.65cm}
\begin{tabular}{lccc}
\hline
Parameter                                          & Model 0S*                     & Model 1S*                    & Model 2S*                     \\ \hline
N$_{\text{H}}$ ($\times\,10^{22}\,\text{cm}^{-2}$) & 13.4$^{\text{+}0.1}_{-0.3}$   & 13.3$^{\text{+}0.2}_{-0.3}$  & 13.3$^{\text{+}0.1}_{-0.2}$   \\
$\Gamma$                                           & 1.78$^{\text{+}0.2}_{-0.3}$   & 1.79$^{\text{+}0.3}_{-0.5}$  & 1.77$^{\text{+}0.3}_{-0.8}$   \\
$f_{\text{SC}}$                                    & 0.04$\,\pm\,0.01$             & 0.03$\,\pm\,0.01$            & 0.03$\,\pm\,0.01$             \\
R$_{\text{in}}$ (R$_{\textsc{isco}}$)              & 2.3$^{\text{+}6.2}_{-1.2}$    & 1.8$^{\text{+}4.2}_{-0.7}$   & 2.45$^{\text{+}7.5}_{-1.0}$   \\
Inclination ($^{\circ}$)                           & 70.4$^{\text{+}9.1}_{-9.2}$   & 74.3$^{\text{+}8.5}_{-12.0}$ & 73.1$^{\text{+}7.5}_{-12.9}$  \\
Spin                                               & 0.90$^{\text{+}0.07}_{-1.31}$  & 0.87$^{\text{+}0.11}_{-1.25}$                         & 0.87$^{\text{+}0.12}_{-1.55}$                          \\
log\,$\xi$                                         & 2.41$^{\text{+}0.15}_{-0.19}$ & 2.47$^{\text{+}0.98}_{-1.49}$ & 2.74$^{\text{+}0.85}_{-0.69}$ \\
Fe/Solar$^*$                                          & 1.14                          & 1.9                          & 2.6                           \\
Corona height (R$_{\text{g}}$)                     & -                             & -                            & 9.7$^{\text{+}8.5}_{-4.0}$    \\ \hline
$\chi^2/\nu/\chi^2_r$                                       & 900/879/1.02                       & 897/879/1.02                      & 898/878/1.02                       \\ \hline
\end{tabular}}
{\begin{flushleft}\small{\bf{Notes:}}  Errors are at 90\% confidence limit. $^*$This parameter was left free but limited to assume values within 1--3.  \end{flushleft}}
\end{table}

\section{Discussion}

We have gathered public available X-ray data from \textit{XMM-Newton}, \textit{NuSTAR} and \textit{INTEGRAL} to build a broadband spectrum (2--200\,keV)  of the black hole candidate \E. Since the observations were not taken contemporaneously, arguments regarding the flux and emission state of the source for each epoch were presented to justify a simultaneous analysis.  We present on the application of several models that describe the three main spectral features of this type of source; the disc, reflection and comptonisation components. The composed spectrum is visually featureless and a relatively good fit ($\chi^2/\nu\,\sim\,1.04$) could already be achieved with a simple disc + powerlaw combination. We have shown, however, that the upgrade to models that account for a reflection feature improves the quality of the fit ($\chi^2/\nu\,\sim\,1.02$); when applicable (i.e., the reflection component was simply added to a combination), F-test values expressed such improvement. As we dealt with many model combinations -- most of which are not nested -- a more sophisticated way to compare the relative quality of fit for different models is using the Akaike Information Criterion (AIC, \citealp{1974ITAC...19..716A}). The AIC for each model $i$ is given by
 
\begin{equation}
\text{AIC}_i\,\text{=\,} \,2\,m - C\, \text{+}\, \chi^2\, \text{+} \frac{2\,m\,\text{(}m\text{+}\,1\text{)}}{n-m-1}, 
\end{equation}
\noindent where $m$ is the number of free parameters, $n$ is the number of energy bins and $C$ is a value that depends only on the data set and thus, is constant for every model. Lower values of AIC indicate better models. Moreover, the probability of a certain model to be the best among a set of models can be computed with the Akaike Weights \citep{aic}, which express the probability that a model $i$ is the best among $k$ models. It is defined as
\begin{equation}
 p_i\,\text{=}\,\frac{\text{exp(}- \Delta \text{AIC}_i/2\text{)}}{\sum_{i\text{=}1}^{k}\,\text{exp(}- \Delta \text{AIC}_i/2\text{)}},
\end{equation}
\noindent with $\sum\,p_k$ = 1 and where $\Delta \text{AIC}_i$ = AIC$_i\, - $\,min(AIC). When computing this to all model combinations we applied, the ones {\em without} a reflection component hold a probability of less than 0.5\% of being the best-fitting models. This means that although the reduced $\chi^2$  among fits differs only in the second decimal place for models with and without reflection, it is very likely that this component -- whether relativistic or not -- is required to properly describe the data. In the following subsections we briefly discuss about the overall resulting values presented throughout the text.

\subsection{The Emission State}

Black hole emission states have been customarily discriminated by the indices of phenomenological powerlaws fitted to their higher energy part of the spectrum; being a steeper index indicative of the high/soft state (HSS) state and a less steep of the LHS.  The powerlaw indices found for our spectra -- individually or combined -- are somewhat steeper than values previously reported for \E. Also different from most reports is the absence of a cutoff energy up to $\sim$\,200\,keV and an atypical higher corona temperature (for some of our models). None the less, with comparable energy coverage from \textit{Suzaku}, \cite{2010AIPC.1248..189R} reported very similar indices ($\Gamma$\,$\sim$\,1.8) and high energy cutoffs to what we found, in epochs they denoted as being right after a transition from the HSS to the LHS. Since there are so far no reports of \E\space being observed in the soft state, we believe this classification was most likely based on the fact that their observations occurred after periods of virtually no detection in the 15--50\,keV band from BAT daily flux measurements. As in our study we simultaneously analyse non-contemporaneous data, other than affirming \E\space was in a ``softer'' LHS 
for the observations, we can not state anything about the source being after or before a transition. Our calculated luminosity (2--200\,keV) of less than a few percent of Eddington's also corroborates with a black hole in the LHS.

\subsection{Disc Requirement and Truncation}

Previous studies with \textit{Suzaku} \citep{2010AIPC.1248..189R} and \textit{NuSTAR} \citep{2014ApJ...780...63N} have reported the need to account for a softer component to fit the spectrum of \E\space down to energies of 2 and 3\,keV, respectively. In fact, significant fit improvement is achieved when we include a disc component -- both when fitting our \textit{NuSTAR} data alone ($>$\,3\,keV, F-test probability of $\sim$\,10$^{-6}$ against no disc) and, even more explicit, when fitting with \textit{XMM-Newton} data included (down to 2\,keV, F-test probability of $\sim$\,10$^{-13}$). This reveals that, despite the very high interstellar absorption towards the source's location, a multitemperature accretion disc is required to describe \E\space spectra above a few keV. 
Also in agreement with the mentioned studies is our estimation of the disc inner radius, which we find -- from the normalisation of the disc component -- to be no more than 60\,R$_{\text{g}}$. When evaluating this same quantity  from modelling the reflection component, our results suggest an even less truncated disc. From performing fits for various reflection models and different inner radii distances from the black hole up to 50\,R$_{\textsc{isco}}$ (roughly 60\,R$_{\text{g}}$ for a Kerr black hole) we notice that the fit quality consistently decreases for increasing radius. When let free, overall best-fits imply the radius is no farther than $\sim$\,15\,R$_{\textsc{isco}}$ ($\sim$\,20\,R$_{\text{g}}$) from the compact object at 3\,$\sigma$ level confidence. Hence, from our results, \E\space may be another evidence to support that discs for black holes in the LHS are not necessarily truncated at very large distances  (e.g., see \citealp{2008ApJ...680..593T} for GX\,339--4, \citealp{2015ApJ...808....9P} for Cyg\,X$-1$ and \citealp{2018ApJ...852L..34X} for MAXI J1535--571).

\subsection{The Reflection Component}

A hint of a skewed iron line and a compton hump in our \textit{NuSTAR} data prompted us to include models to describe these reflection features which, as previously pointed out, improved the quality of the fit -- endorsing its presence in the spectrum. Based on the possibility of the disc inner radius being close to the black hole, we advanced to relativistic reflection models, whose main parameters of interest are the spin of the black hole and the inclination of the disc.
All model combinations indicate a high spin and a high inclination. Every best fit sets the parameter spin to values close to that of a maximum speed rotating black hole (a$_*$ = 0.998), some with 1\,$\sigma$ confidence that a$_*\,\gtrsim$\,0.5. Reports of high spin values for black holes in binary systems are very frequent, particularly when the estimate procedure used is the iron line method (see, e.g., Table 2 from \citealp{2018AnP...53000430B}).  None the less, despite this trend, our assessment is that our models could not adequately constrain the spin.
As for the disc inclination, best-fitting values vary roughly from  60$^{\circ}$ to almost 80$^{\circ}$ among models but they all agree, within 3\,$\sigma$, that it is $\gtrsim$\,50$^{\circ}$. These inclinations are consistent with the bi-polar radio jets presence reported by \cite{1992Natur.358..215M}. More recently, \cite{2015A&A...584A.122L} obtained constraints on the jets inclination, whose values -- if assumed the disc-jet perpendicularity -- are also consistent with fairly high inclination for the accretion disc. Such high inclinations would suggest that detectable eclipses should occur. However, in both reports on the orbital period ($\sim$12.6 days) of \E\space -- from long--term time analysis in different X-ray bands (2.5--12.5\,keV, \citealp{2002ApJ...578L.129S}; and 15--50\,keV, \citealp{2017ApJ...843L..10S}) -- no obvious shape of an eclipse was identified. This could be due to an extended hard X-ray emitting source that is only partially obscured by the companion star. With an extended corona, either in length or height, photons reprocessed in the disc could be subject to being upscattered in the corona again -- a scenario that would be in agreement with the not very pronounced reflection feature in our spectrum and with the small reflection fraction values we find (see, e.g., \citealp{2015MNRAS.448..703W}).

\subsection{The Mass}

A virtually identical fit quality can be achieved when the simple multitemperature \texttt{diskbb} model -- parametrised only by its inner temperature -- is replaced by \texttt{kerrbb}, a much more complex model dependent on several parameters.  
Different choices for the spectral hardening value have not caused significant changes in the mass, spin, accretion rate or inclination. This weak relationship between $f$ and these parameters had been demonstrated by \cite{1995ApJ...445..780S} and also, e.g., recently verified (\citealp{2019MNRAS.487.4221S}).
The extremely low accretion rates such as the ones we find (equivalent to roughly $\sim$\,10$^{-11}$\,M$_{\odot}$\,$\cdot$\,yr$^{-1}$) had been reported for black hole binaries in the ``quiescent'' state (see, e.g., \citealp{2004A&A...421...13P} and references therein). \cite{1998tbha.conf..148N} argue that for such accretion inefficiencies, the thin disc approximation is no longer valid and that the inner part of the disc would be better described by an advection-dominated accretion flow model; this issue is, however, beyond the scope of this paper.
Finally, overall results suggest a relatively low black hole mass in \E, with a median best-fitting value of 4.7\,M$_{\odot}$ amongst all 18 model combinations; in only 2 combinations a mass up to 10\,M$_{\odot}$ is allowed at 90\% confidence level.
An inspection of the BlackCAT \citep{2016A&A...587A..61C}, a catalogue for stellar-mass black holes in X-ray transients, shows that -- at its time of publication -- 18 low mass X-ray binaries (LMXBs) had the black hole mass dynamically measured (i.e. from the companion's radial velocity measurement method). Considering their uncertainties, roughly all black hole masses lie somewhere within a 5\,M$_{\odot}$\,$\lesssim$\,M\,$\lesssim$\,12\,M$_{\odot}$ range. Additional research in the literature adds 6 more LMXBs with mass estimates to the list, of which 4 were determined by different methods based on X-ray spectra (10\,$\pm$\,0.1\,M$_{\odot}$ for 4U\,1630$-$47,  \citealp{2014ApJ...789...57S}; 4.7--7.8\,M$_{\odot}$ for MAXI\,J1659$-$152, \citealp{2016MNRAS.460.3163M}; 10.31--14.07\,M$_{\odot}$ for H\,1743$-$322, \citealp{2017MNRAS.466.1372B} and 10.62--12.33\,M$_{\odot}$ for IGR\,J17091$-$3624, \citealp{2018arXiv180805556R}) -- as their optical/infrared counterparts have not yet been confirmed . The reported masses also lie in the same range of those LMXBs from the BlackCAT. The situation for high mass X-ray binaries (HMXBs) is more modest, as only 3 black holes have their masses measured: Cyg\,X$-1$, $14.8\,{\pm}\,1$\,M$_{\odot}$; LMC\,X$-$1, $10.9\,{\pm}\,1.4$\,M$_{\odot}$ (see, e.g., \citealp{2014SSRv..183..223C}) and LMC\,X$-$3, $6.98\,{\pm}\,0.56$\,M$_{\odot}$ \citep{2014ApJ...794..154O}.  As for neutron stars, no more than a couple of objects have been reported to surpass 2\,M$_{\odot}$ (e.g., \citealp{2018ApJ...859...54L}, \citealp{2019NatAs.tmp..439C}). These numbers indicate an absence of compact objects around the 2--5\,M$_{\odot}$ mass range, referred for the first time by \cite{1998ApJ...499..367B} as the \textit{mass gap}. Thereby, if the estimate presented here of a black hole mass around 4--5\,M$_{\odot}$ (i.e., considering the median value from all best fits) is further confirmed, \E\, would be one of the first compact objects to populate this gap. 
It should be pointed out that -- due to their positive correlation -- by fixing the accretion rate to higher values (e.g., the initial guess) the black hole mass increases accordingly; yet, no statistically acceptable fit could be achieved in these circumstances.

\section{Conclusions}

We have presented and discussed the application of several model combinations in order to fit a broadband X-ray spectrum (2--200\,keV, built from \textit{XMM-Newton}, \textit{NuSTAR} and \textit{INTEGRAL} data) of the black hole candidate \E, system in which many dynamical parameters are still unknown. Models were applied in gradual increase order of complexity; significant enhancements in the quality of the fit are accomplished when a thermal disc and, further, a reflection component are included -- expressing that such components are needed to describe the spectrum. The main conclusions regarding the results from modelling the spectrum are summarised below.

\begin{enumerate}
 
 \item \E\space was in a very similar emission state during all three observations, namely the LHS. Although slightly steeper than previously reported for \E, the powerlaw indices found -- either for the individual spectra (Table \ref{tab:01}) or for the composed spectrum (Table \ref{tab:02}) -- are still covered by the accepted index range for black holes in the LHS. The Eddington's luminosity calculated and presented for the composed spectrum is also in agreement with a black hole in this state.  
 
 \item Estimates of the inner disc radius -- whether calculated from the disc normalisation ($\lesssim$\,60\,R$_{\text{g}}$) or from modelling the reflection component ($\lesssim$\,20\,R$_{\text{g}}$) -- indicate an inner disc not far from the compact object; this is somewhat conflicting with the scenario of very truncated accretion discs, widely regarded to be the case for sources in the LHS.
 
 \item Computing the Akaike Weights for every model combination we applied shows that the ones that include a reflection component are 99.5\% more likely to be the best-fitting models. Results from modelling this component point to a high disc inclination of at least 50$^{\circ}$ (3\,$\sigma$), which is in agreement with previous studies in radio. The spin parameter, for every best fit, is set to values close to a$_*$ = 0.998; however, it is only constrained to be a$_*\,\gtrsim$\,0.5 at 1\,$\sigma$.  
 
 \item Modelling the low energy part of the spectrum with the mass-dependent disc model \texttt{kerrbb} provided black hole mass values with no more than 10\,M$_{\odot}$. Actually, in only 2 out of 18 model combinations the resulting mass could be as high as this value; the median best-fitting mass value amongst all combinations is only 4.7\,M$_{\odot}$. If such low mass is further confirmed, the black hole in \E\space would be amongst one of the first compact objects to populate the so-called 2--5\,M$_{\odot}$\,\textit{mass gap}.

\end{enumerate}

\section*{Acknowledgements}

P.E. Stecchini and M. Castro acknowledge FAPESP for financial support under grants \#2017/13551-6 and \#2015/25972-0, respectively. J. Braga also acknowledges FAPESP for support under Projeto Tematico \#2013/26258-4.  The authors thank the referee for detailed and constructive comments that helped us improve the paper.








%
%


\bsp	
\label{lastpage}
\end{document}